\documentclass[aps,pre,twocolumn,superscriptaddress,showkeys,showpacs]{revtex4-2}

\usepackage{amsmath,graphicx,amssymb,hyperref}
\usepackage{tabularx,array,color} 
\usepackage{placeins}
\newcolumntype{P}[1]{>{\centering\arraybackslash}p{#1}}

\usepackage[margin=1.5cm]{geometry}
\usepackage{tabularray}
\usepackage[tiny, center, uppercase]{titlesec}

\usepackage[title]{appendix}

\begin{document}

\title{Structure of road networks and \\the shape of the macroscopic fundamental diagram}

\author{Erwan Taillanter}
\email{erwan.taillanter@ipht.fr}
\affiliation{Universit\'e Paris-Saclay, CNRS, CEA, Institut de Physique Th\'{e}orique, 91191, 
Gif-sur-Yvette, France}

\author{Andreas Schadschneider}
\email{as@thp.uni-koeln.de}
\affiliation{Institut für Theoretische Physik, Universität zu Köln, 50937 K\"oln, Germany}

\author{Marc Barthelemy}
\email{marc.barthelemy@ipht.fr}
\affiliation{Universit\'e Paris-Saclay, CNRS, CEA, Institut de Physique Th\'{e}orique, 91191, 
	Gif-sur-Yvette, France}
\affiliation{Centre d'Analyse et de Math\'ematique Sociales (CNRS/EHESS) 54 Avenue de Raspail, 75006 Paris, France}

\begin{abstract}

The macroscopic fundamental diagram (MFD) is a large scale description of the traffic in a urban area and relates the average car flow to the average car density. This MFD has been observed empirically in several cities but how its properties are related to the structure of the road network has remained unclear so far. The MFD displays in general a maximum flow $q^*$ for an optimal car density $k^*$ which are crucial quantities for practical applications. Here, using numerical modeling and dimensional arguments, we propose scaling laws for these quantities $q^*$ and $k^*$ in terms of the road density, the intersection density, the average car size and the maximum velocity. This framework is able to explain the scaling observed empirically for several cities in the world, such as the scaling of $k^*$ with the road density, the relation between $q^*$ and $k^*$ and the impact of buses on the overall capacity $q^*$. This work opens the way to a better understanding of the traffic on a road network at a large urban scale. 

\end{abstract}


\maketitle


\section{Introduction}

\begin{figure}
  \includegraphics[width=1.0\linewidth]{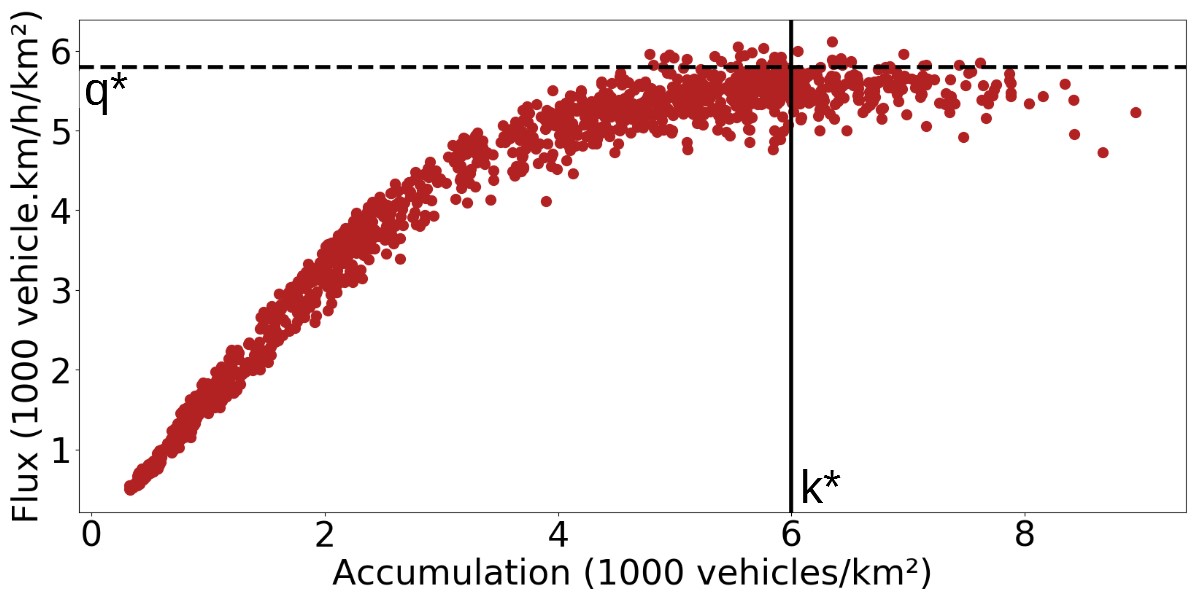}
  \makeatletter\long\def\@ifdim#1#2#3{#2}\makeatother
    \caption{Macroscopic Fundamental Diagram of Paris (France). Each point represents the traffic state of the network at a given date. The vertical line indicates the  optimal density $k^*$ and the horizontal dashed line the optimal flow $q^*$. This MFD was obtained for the period April-May 2014 (data from \href{https://opendata.paris.fr/explore/dataset/comptages-routiers-permanents/export/?disjunctive.libelle&disjunctive.etat_trafic&disjunctive.libelle_nd_amont&disjunctive.libelle_nd_aval}{\texttt{opendata.paris}})}
    \label{MFD}
    \vspace{-3mm}
\end{figure}

Traffic congestion in large urban areas is a crucial problem as a source of pollution and loss of time. Congestion is not only a product of insufficient road capacity \cite{DurantonT11} and it is clear now that the structure of the road network has to be carefully planned in order to improve the traffic. 
A crucial tool for monitoring the traffic at a urban scale is the Macroscopic Fundamental Diagram (MFD) that relates the traffic flow and the density of a (large) road network based on highly aggregated data. It extends to networks the concept of the fundamental diagram - relating the flow to the density of vehicles - , discovered by Greenshield almost a century ago \cite{Greenshield1933}, which is arguably the most important quantitative characterization of traffic for single roads. Although the concept of a MFD, describing the response of the network to traffic demand, was already known in the 1960s, its full potential has only been realized with the work of Daganzo and Geroliminis \cite{Daganzo07,GeroliminisD08}. These authors were the first to study the relationship between the average density (also called \emph{accumulation}) and the average traffic flow at the scale of a urban road network, coining the term Macroscopic Fundamental Diagram. Their empirical discovery that both average quantities are related by a well defined relationship (see Fig.~\ref{MFD} for an example), similarly to the single road case, has been a major step in our comprehension of traffic in urban areas. Much like the Fundamental Diagram, the MFD displays a maximum $q^*$ for a critical density of cars $k^*$.
Subsequent work focused on understanding the origins of such a network-wide relation \cite{DaganzoG08, GeroliminisS11, Qi22apre, LeclercqM18, KnoopH13, KnoopLH15} and on the characterization of its functional form \cite{Loderetal19, AmbuehlLBMA20, LeclercqCT14, JohariKLNM21}. This diagram has important applications such as \emph{perimeter control} \cite{WuLJHH18, GuoYHG21}, which tries to maintain an optimal network performance at maximal flow, \emph{congestion pricing} \cite{GuSLS18} and its role in network-wide congestion propagation just starts to be understood \cite{AmbuhlMG23}.

In general, the MFD depends on characteristics like traffic demand, the structure of the road network and the fundamental diagrams of the constituting roads as well as other factors (heterogeneity of traffic, driver behavior as route choice etc.). However, the question of the relation between the network and the MFD remains unclear \cite{KnoopJH14, LoderAMA19}. Finding the parameters of the network which influence the MFD is crucial for taking the best decisions in future road planning. In this respect, the work \cite{LoderAMA19} is an important step for our understanding, as the authors propose an empirical study of the MFD using billions of vehicle observations for more than $40$ cities worldwide and tried to correlate the MFD with network properties. In particular, they found a sublinear relation between the critical accumulation $k^*$ and the network density (given by the length of roads per unit area). 



In this article, we focus on the problem of the relation between the critical parameters $q^*$ and $k^*$  of the MFD (which define in some way its shape), and the properties of the network. In order to investigate this problem, our main tool will be numerical simulations of the traffic on a network, and vary its parameters independently. These numerical results enable us to propose and to test scaling laws for these critical parameters $q^*$ and $k^*$ in terms of the network parameters.

\section{Simulations}

\subsection{The model}

We simulated the traffic with the ChSch model \cite{ChowdhuryS99} which was introduced as an extension to urban road networks of the NaSch model \cite{NaSchmodel} for highway traffic. It allows to study the interplay between vehicle dynamics and interactions induced by intersections controlled by traffic lights \cite{BrockfeldBSS01, GaoLLHMZ07}. 
We consider a regular 2D-lattice of roads and for the sake of simplicity, we consider cars moving from West to East and from South to North only. On each road, the cars move by following the rules of the ChSch model that we describe briefly (for more details on this models, see for example \cite{schadschneider2010stochastic}). The road is divided into discrete cells and time is also discretized. Each cell can be either empty or occupied by a vehicle with a velocity $v$ that can take integer values in $[0,v_{max}]$ where $v_{max}$ is the maximum speed allowed on the road. The update is performed in parallel according to some rules for acceleration and braking. Contrary to the original paper, we do not introduce a probability for cars to randomly decelerate. At each time step $\tau$, the speed $v_i$ of each car is updated in parallel following, if the traffic light in front is green: 
$$v_i(t+\tau) = \textrm{max}(v_{max}, v_i(t)+1, d_{i,i+1}(t)-1)$$
where $v_i$ is the speed of the car $i$, $v_{max}=5$ cells/timestep and $d_{i,i+1}$ is the number of cells separating the car $i$ from its predecessor. 
If the traffic light in front of car $i$ is red, this rule changes to:
$$v_i(t+\tau) = \textrm{max}(v_{max}, v_i(t)+1, d_{i,i+1}(t)-1, d_{i,tl}(t))$$
where $d_{i,tl}(t)$ is the distance to the traffic light. In particular, note that the (i) the speed of the car $i$ only depends on the position of car $i+1$, not on its speed, and (ii) cars accelerate at most by 1 cell/$\tau^2$, but can decelerate by up to $5$ cells/$\tau^2$ if needed. Once all the speeds are determined, the cars progress together by a discrete number of simulation cells, the new positions of car $i$ being:
$$d_i(t+\tau)=d_i(t)+v_i(t+\tau)$$

\begin{figure}
    \centering
    \begin{minipage}{\linewidth}
    \includegraphics[width=1.1\linewidth]{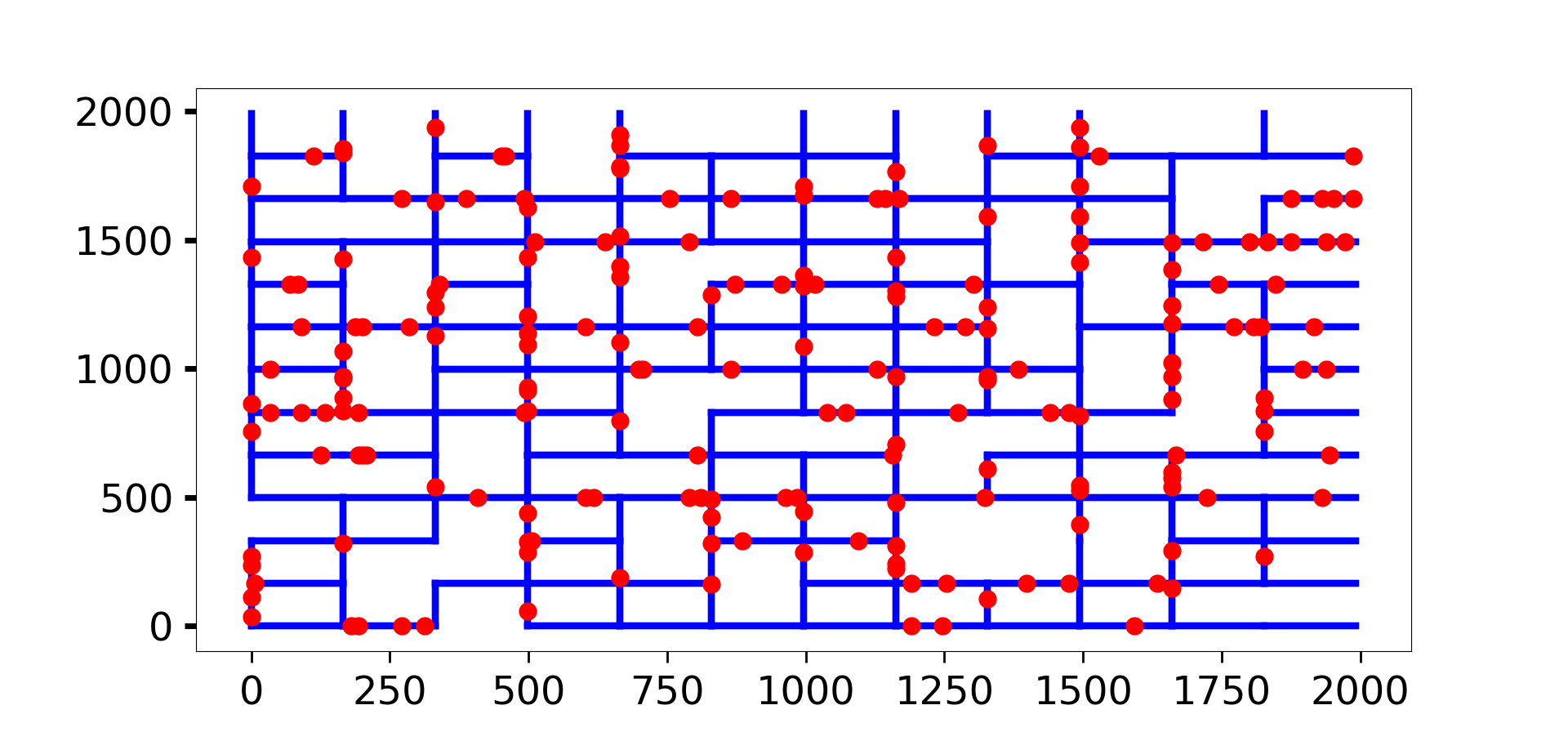}
    \end{minipage}
    \begin{minipage}{\linewidth}
    \includegraphics[width=1.1\linewidth]{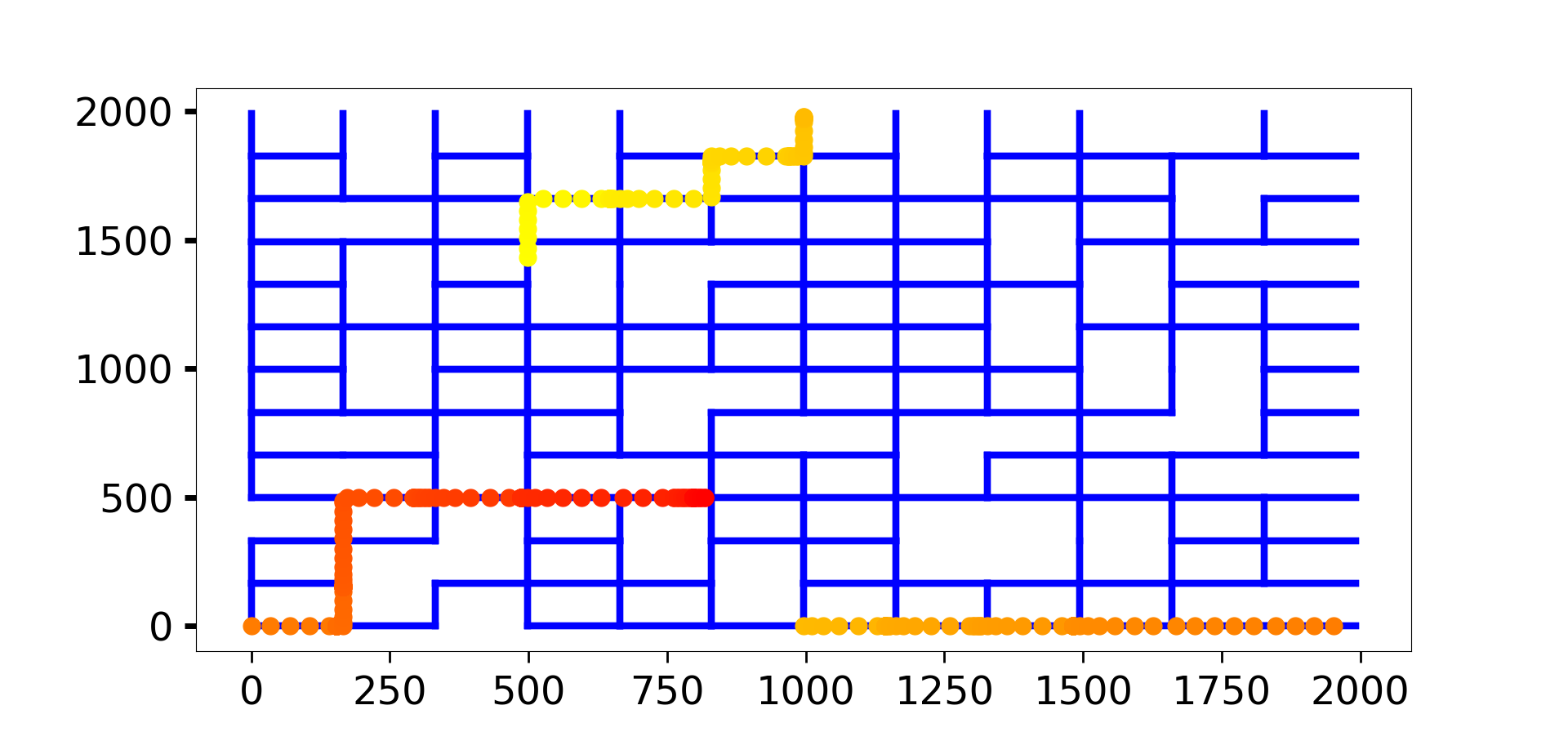}
    \end{minipage}
    \begin{minipage}{\linewidth}
    \includegraphics[width=1.1\linewidth]{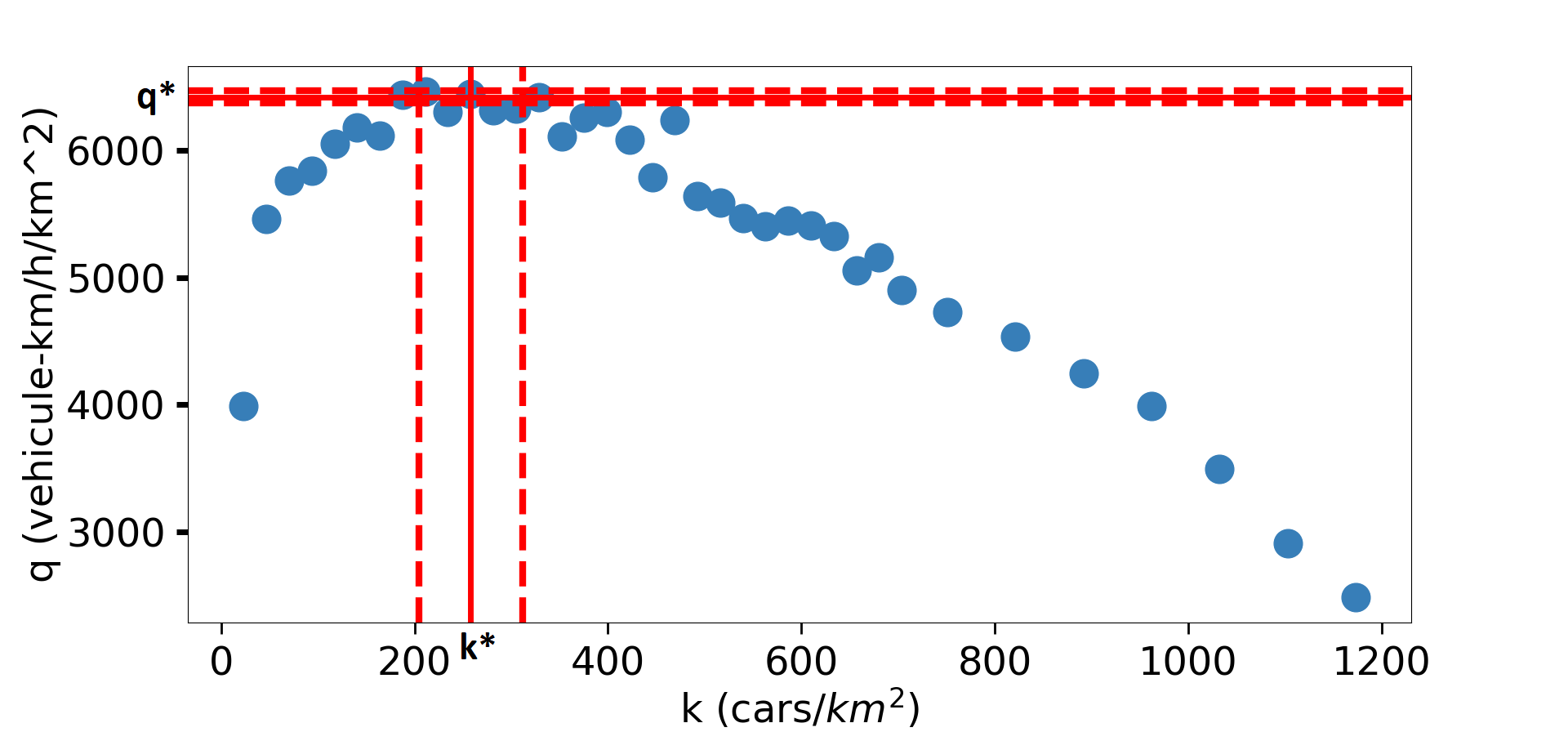}
    \end{minipage}
    \caption{Illustration of our simulation for a network with $p=0.2$, with $13\times 13$ nodes and roads of length $166m$ (distances are indicated in meters). We use periodic boundary conditions. (Top) Initial time: $N=200$ cars located at random in the network. (Middle) Trajectory of a car during $100$ consecutive timesteps (time goes from yellow to read). (Bottom) Corresponding MFD obtained over $500$ timesteps. We show the values of $k^*$ and $q^*$ with the corresponding confidence intervals (red dashed lines). The shape of the MFD implies in general that the estimate of $k^*$ is noisier than the one for $q^*$.}
    \label{simulation}
\end{figure}

In addition and in contrast with the original model \cite{ChowdhuryS99}, at each intersection, every car picks a new direction (east or north) at random and continues its progression. Road networks are, however, not regular lattices (see for example \cite{barthelemyspatial} and references therein), and we introduce a probability $p$ for each link of the lattice to be removed. Starting from a regular 2D-lattice, we can transform the network into a less regular network, with intersections of degree $3$ and $4$ (see figure \ref{simulation}).

In the spirit of Loder et al. \cite{LoderAMA19}, we define all our variables as averages per unit area: $k^*$ is in veh/km$^2$, and $q^*$ in cars$\cdot$km/h/km$^2$. The network is defined by the fraction $p$ of missing links, the density $\rho_r$ of roads in space (with unit km/km$^2$) and the density  $\rho_i$ of intersections in space (with unit km$^{-2}$). The progression of cars using the ChSch model is defined by 4 parameters: the length $L_{car}$ (km) of the discrete cells on which cars evolve, the time step on which we update the speeds and positions (taken equal to 2 seconds), the maximal speed of the cars $v_{\max}$ (expressed in our model as a number of steps per time step) and the duration of each green or red light phase, $t_{tl}$ (in seconds).

\subsection{Measuring the MFD}

\begin{figure}
    \centering
    \includegraphics[width=\linewidth]{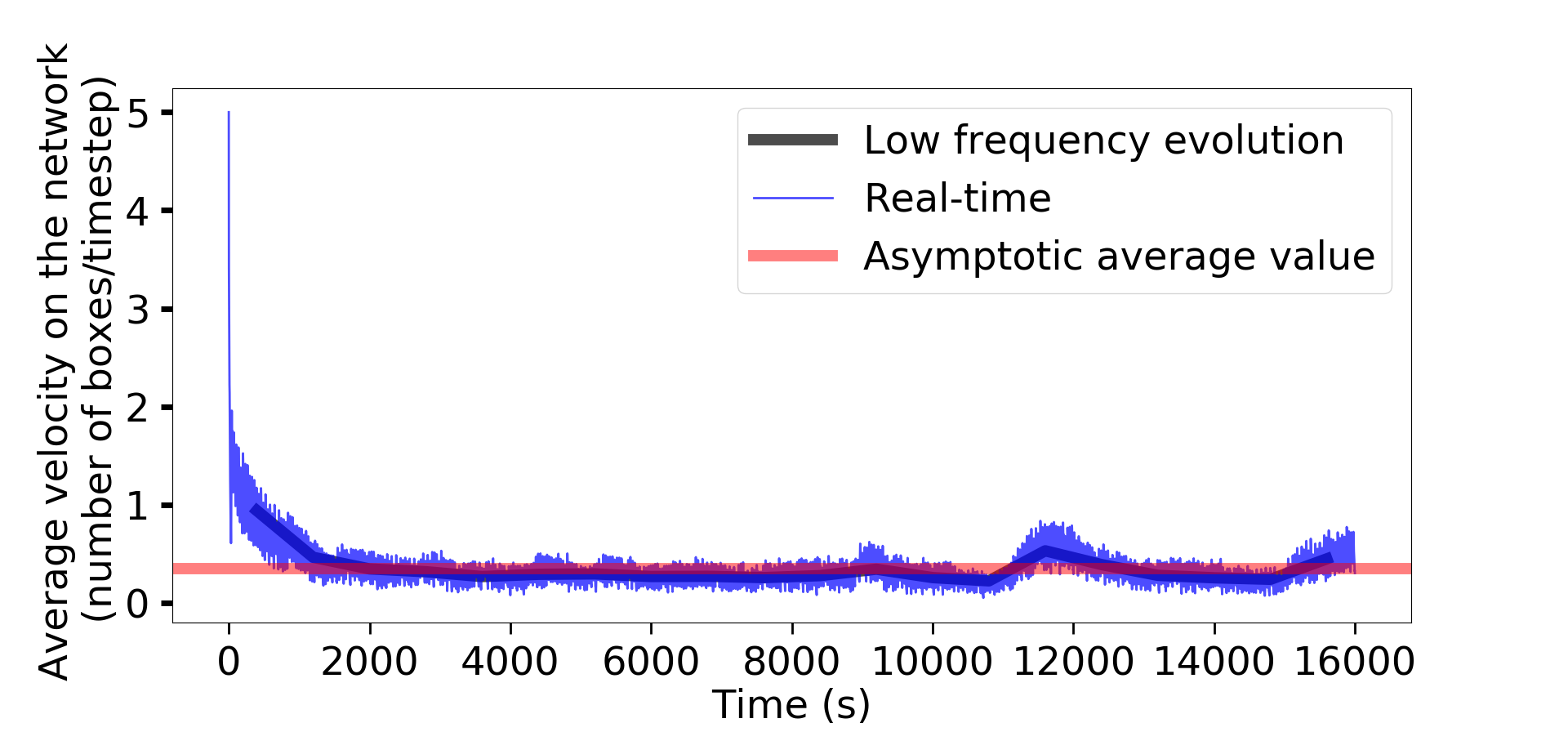}
    \caption{Evolution of the average velocity on the network with time. The density on the network is here close to $k^*$. Starting from a random configuration, the average speed decreases as traffic jams emerge during the first $1000$ seconds. The network then remains in a steady state with relatively constant average velocity.}
    \label{fig:velocity_evolution}
\end{figure}

Each point of the MFD is obtained by averaging the density of cars and the flux on the network over a given time span. The choice of this time span is important in real-world network. It should be longer than the typical duration of green light cycles, as we are interested in the average state of traffic rather than in its high frequency variations. On the other hand, averaging over too long a period means that some cars will enter or exit the network, resulting in a biased estimation of the state of the traffic. This problem is also well-known for the road fundamental diagrams, where the measured diagram is always smaller than the `real' diagram. The way we obtain the MFD is then the following:
\begin{enumerate}
    \item We pick a network corresponding to the desired values of the parameters
    \item We pick a number of cars and add them randomly to the network. This stage determines the car density $k$ (ranging from $1\%$ to $50\%$ of the maximal number of cars one could fit on the network, i.e. values of density respectively far smaller and larger than the MFDs maximum).
    \item We let the network evolve during $500$ time steps towards its equilibrium
    \item We let the network evolve during $500$ times steps and estimate the average flow $q$ during this period
    \item We repeat steps 3 and 4 for the same network but an increasing number of cars.
\end{enumerate}
Hence, for each estimation of the flux, the density $k$ is fixed (unlike for real life measurements). This allows us to estimate the flow over a longer period of time than what would typically be done on real networks. Step 3 is necessary to have a realistic MFD. Indeed, on real networks, cars are never randomly distributed. We determined the time needed to reach the equilibrium using the average velocity. Starting from our random initial configuration, we measure the average velocity on the network as a function of time. We observe a decrease of the velocity as cars regroup in traffic jams in specific points of the network. The choice of 500 time steps (or $1,000$ seconds) corresponds to a duration which guarantees that the network has reached a steady state (average velocity is constant, see figure \ref{fig:velocity_evolution}).

Step 4 also needs some clarification. The average flow can be estimated with two methods, which are subject to debate amongst traffic engineers. Remembering that the average flow is expressed as a number of travelled kilometers during a period of time and per unit surface, one can either sum the speeds of the cars at each time step, average this sum over all time steps and finally divide by the network surface. Alternatively, one can measure the number of cars exiting a road at each time step and consider that this car covered the length of the road during that time step. Summing these distances over all time steps, dividing by the duration of the measurement (here $500$ time steps correspond to $1,000$ seconds) and dividing by the network surface leads to another estimation of the flux. The former method corresponds to real life data provided by onboard speed measurements (Floating Car Data, FCD), while the latter corresponds to measurements obtained with fixed loop detectors on the road (Loop Detector Data, LDD). Whilst one might intuitively think that both methods are equivalent, they are not and the resulting MFD depends - albeit slightly - on the method chosen. The choice of a method, or a mix between both of them is not settled among the traffic engineer community, but here we emulate LDD results, mainly because we wanted to be as close as possible to the conditions of the empirical measures of Loder et al. \cite{LoderAMA19}.

We use the network parameters described in Table \ref{tab:Table}. In real networks, $\rho_r$ and $\rho_i$ can vary independently, thanks to multilane roads, the sinuosity of roads, as well as the existence of bridges and tunnels (i.e. roads might physically cross without intersecting). In order to reproduce this in our simulation, we introduce a geometrical factor $g$ linking the actual length of the roads $\L_{road}$ to the geometrical distance between intersections $\L_{grid}$, such that $\L_{road} = g \L_{grid}$.

\begin{table*}
\begin{tabular}{ | P{1.5cm} | m{10cm}| m{2.5cm} | P{2.5cm} | }
  \hline
  \textbf{Variable} & \textbf{Description} & \textbf{Unit} & \textbf{Default value} \\   \hline
  $q^*$ & Network capacity & $km/h/km^2$ & \\ 
  \hline
  $k^*$ & Critical accumulation & cars$/km^2$ & \\   \hline
  $\rho_r$ & Density of roads in space & lane$\cdot km/km^2$ & \\   \hline
  $\rho_i$ & Density of intersections in space & $km^{-2}$ & \\ \hline
  $p$ & Fraction of links missing compared to regular lattice & Number $\in[0;1]$ &0 \\ \hline
  $L_{car}$ & Length of our simulation cell & $km$ & $7m$ \\ \hline
  $\tau$ & Discrete timestep of the simulation & $h$ & $2$ seconds \\ \hline
  $v_{max}$ & Maximal speed on our simulated network (in cells per timestep) & $h^{-1}$ & $5$ cells/$\tau$ \\ \hline
  $t_{tl}$ & Traffic light timing (duration of a green or red light) & $h$ & $30$ seconds \\ \hline
  $v_s$ & Average speed of the cars in our simulation, in cells per timestep & $h^{-1}$\\ \hline
  $V_{lim}$ & Average speed on the empty network, combination of $L_{car}$, $v_{max}$ and $t_{tl}$ & $km/h$ \\ \hline
\end{tabular}
\caption{Definition of the variables used in this article.}
 \label{tab:Table}
\end{table*}

We then have $\rho_r=\rho_r^0(1-p)=\frac{2g(1-p)}{L_{grid}}$ and $\rho_i=\frac{1}{L_{grid}^2}$, for the density of roads and intersection, respectively, which we can indeed vary independently. Note that the influence of $p$ is twofold : there is an implicit dependence in $\rho_r$ and an explicit one that we explored at fixed $\rho_r $.

We note that, by design, cars travel an integer number of cells per timestep. This means that, when we vary the length of the cars, we also vary their speed. More precisely $V = 2 V_s L_{car}$ with $V$ the real speed ($m/s$), as used in the article, and $V_s$ the simulation speed (cells/timestep). In this article, we plot the data collapse $q^*L_{car}/\rho_r$ versus $\rho_r/L_{car}\rho_i$. To achieve this in practice, we actually plot $q^*/\rho_r$ versus $\rho_r/L_{car}\rho_i$.

\section{Scaling for $q^*$ and $k^*$}

Despite its simplicity, this model is complex enough to display a MFD, from which we extract $q^*$ and $k^*$ (see Fig.~\ref{simulation}). We compute those critical parameters for various values of $\rho_r$, $\rho_i$, $p$, $t_{tl}$ and $L_{car}$, on a $13\times13$ nodes network, running $100$ simulations for each set of values to estimate the error. This is quite
a small network when compared to real-world cases. This choice is motivated primarily by computational resources, as the computation time scales as the square of the number of cars on the network. Besides, because we chose periodic boundary conditions, we do not expect the behavior of the simulation to be radically different on larger networks. We tested this claim on a subset of the results presented here, and found that the results are very similar on a $20\times20$ nodes network, when compared to a $13\times13$ nodes network (see Appendix A).

\begin{figure}[ht!]
    \includegraphics[width=\linewidth]{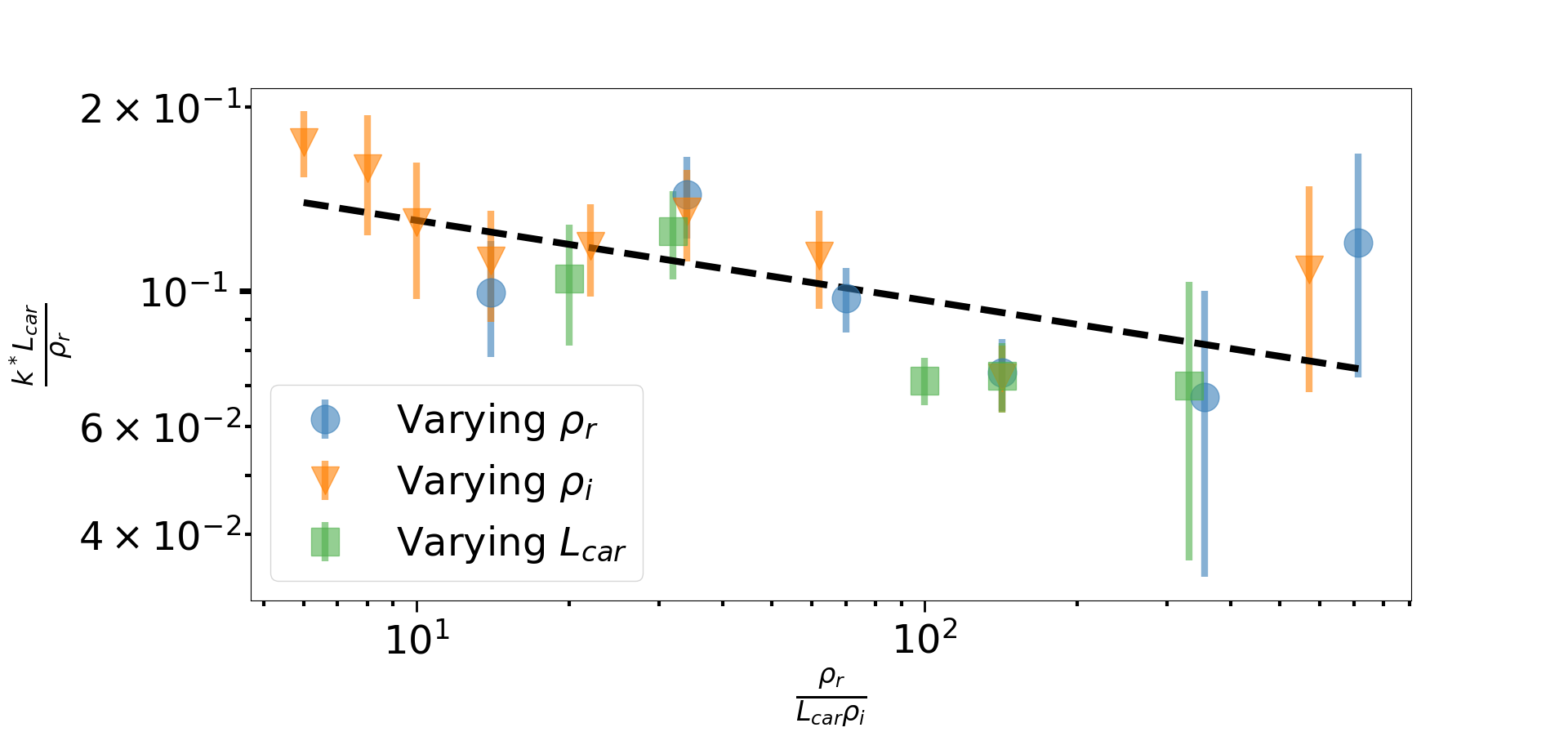}\\
    \includegraphics[width=\linewidth]{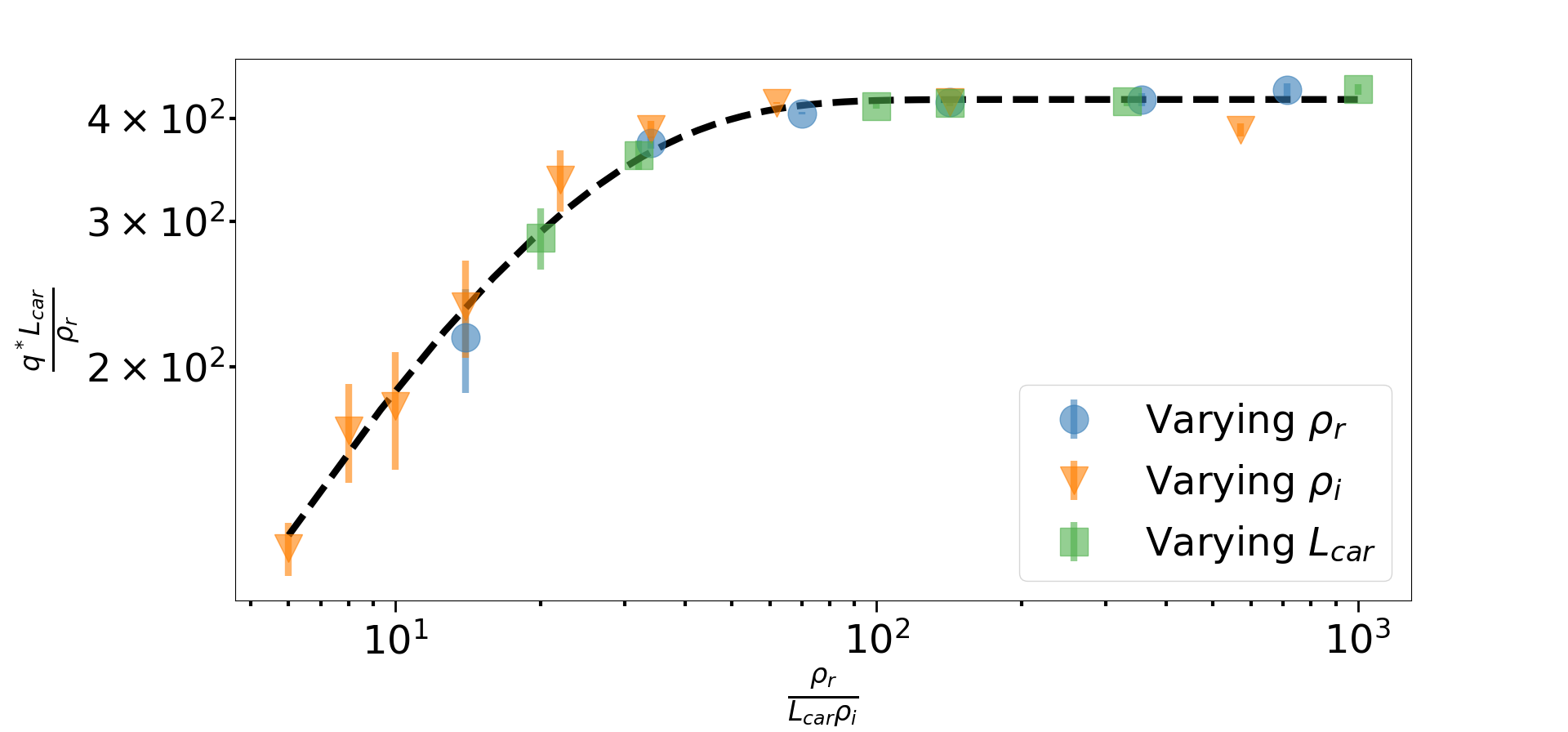}
    \caption{Test of the scaling ans{\"a}tze Eqs.~(\ref{eq:scaling1},\ref{eq:scaling2}). We plot the rescaled variables $k^* L_{car}/\rho_r$ and $q^*/\rho_r$ versus the adimensional variable $n=\rho_r/(L_{car}\rho_i)$.
    In both cases we observe a good data collapse. (Top) For $k^*$, we observe a roughly constant function (the dashed line is a power law fit with exponent $\approx -0.1$, $R^2 \approx 0.5$). (Bottom) For $q^*$, we observe an increasing function followed by a plateau, which we fit with an exponential function ($R^2=0.98$). All figures where obtained using 100 random configurations of a network of 13x13 nodes, with $p=0$.}
    \label{data collapse}
\end{figure}

We will use these numerical simulations to identify scaling laws for the parameters $k^*$ and $q^*$. We start with the most general form for these quantities that are
\begin{align}
k^*&=F(\rho_i,\rho_r,L_{car},v_{\max},p,...)\\
q^*&=G(\rho_i,\rho_r,L_{car},v_{\max},p,...)
\end{align}
where $F$ and $G$ are unknown functions at this point. The critical accumulation $k^*$ has the unit of $\rho_r/L_{car}$ while the $q^*$ has the unit of $k^*$ times a velocity. We first define the adimensional quantity $n = \rho_r/L_{car}\rho_i$ which is the average number of cars which can be fitted on a road between two intersections of the network.
Keeping the number of relevant parameters minimal, the natural scaling for $k^*$ and $q^*$ can then be written as
\begin{align}
k^*&=\frac{\rho_r}{L_{car}} \Phi\left(n,p,t_{tl},v_{\max}\right)
\label{eq:scaling1}
\\
q^*&=\frac{\rho_r}{L_{car}} V^*\left(n,p,t_{tl},v_{\max}\right)
\label{eq:scaling2}
\end{align}
where the functions $\Phi$ and $V^*$ are at this stage also unknown. $\Phi$ has no dimension, while $V^*$ is the average speed on the network at the critical point. We tested numerically these scaling forms Eqs.~(\ref{eq:scaling1}), (\ref{eq:scaling2}) by plotting $k^* L_{car}/\rho_r$ and $q^*L_{car}/\rho_r$ versus $n=\rho_r/L_{car}\rho_i$ (at fixed $p$ and $t_{tl}$). If they are correct, we should obtain a data collapse for various values of $\rho_r$, $\rho_i$ and $L_{car}$. We show in Fig.~\ref{data collapse} this data collapse (for $p=0$) which confirms that our scaling assumptions are correct.
A power law fit of the form $\Phi(n) \sim n^\beta$ gives the following estimate (obtained by multiple fits on all the parameters, see Appendix B)
\begin{equation}
    \beta \approx -0.1 \pm 0.1
    \label{eq:beta}
\end{equation}
with $R^2\approx 0.5$. The small exponent combined with the quality of the fit indicates that at first order, we should consider the influence of $n = \frac{\rho_r}{L_{car}\rho_i}$ on $k^*$ to be minimal.

\begin{figure}[h!]
    \includegraphics[width=\linewidth]{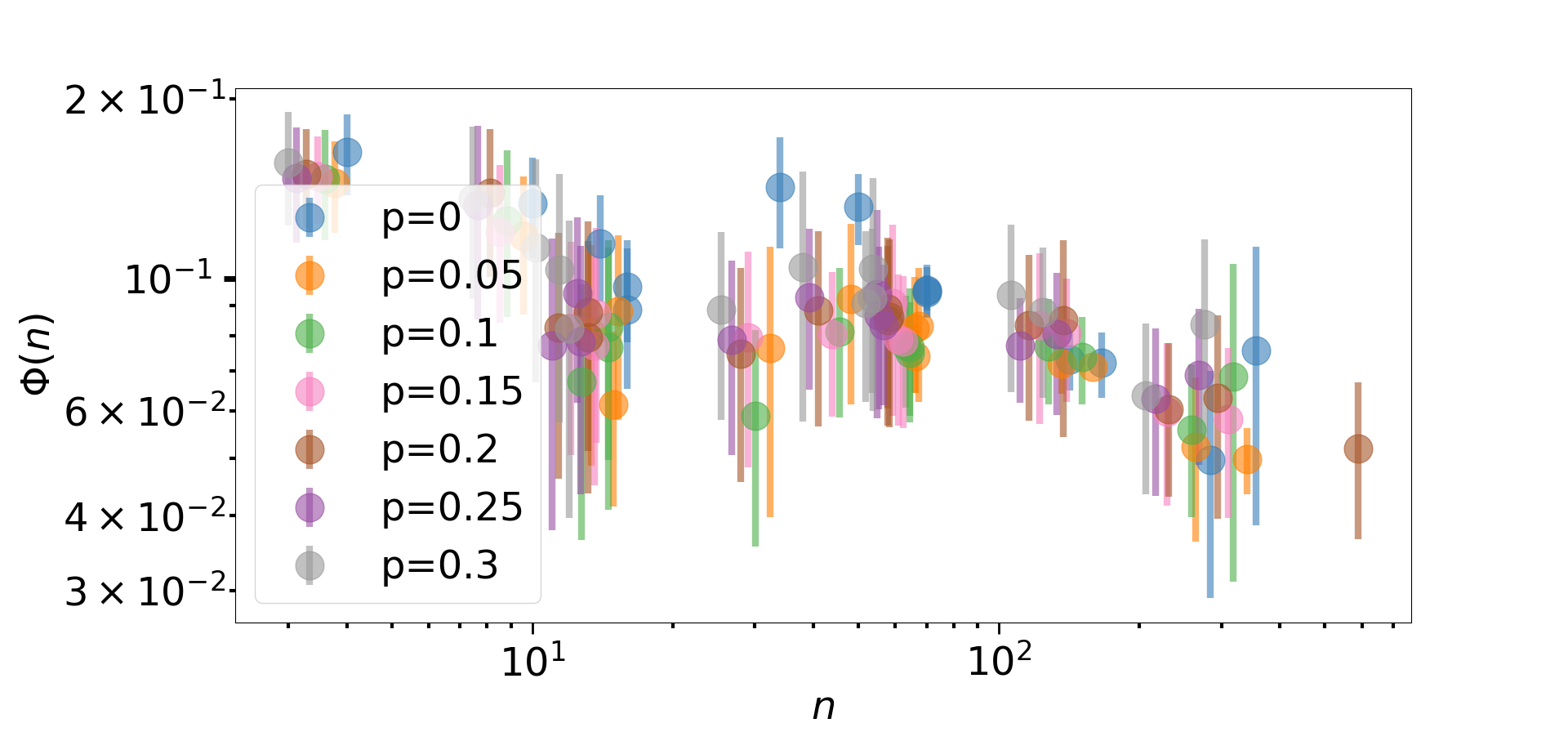}\\
    \includegraphics[width=\linewidth]{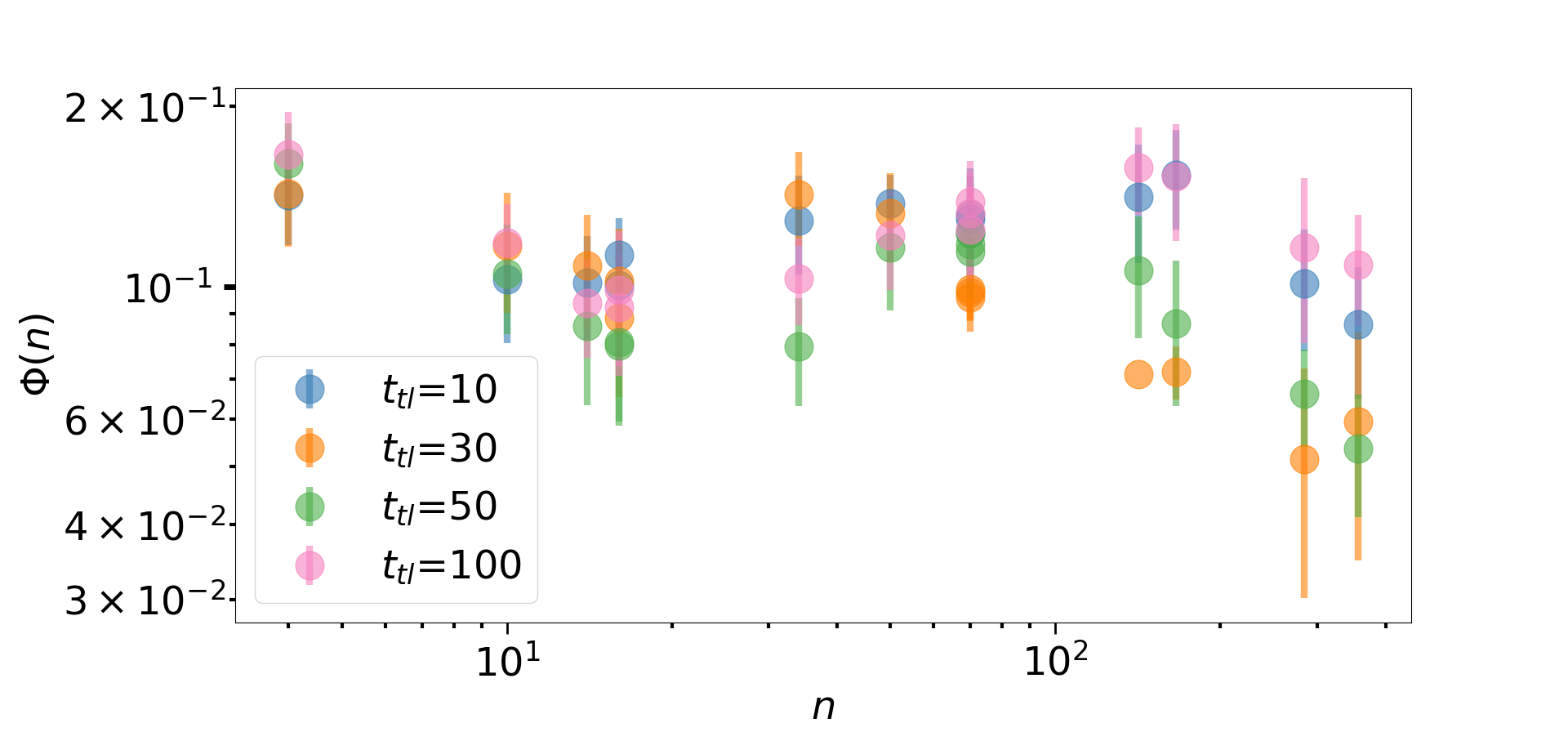}
    \caption{Data collapse for $k^*$ for different values of $p$ (at fixed $t_{tl}=30$ seconds) and $t_{tl}$ (at fixed $p=0$). Neither $p$ nor $t_{tl}$ have a significant influence on $\Phi$}
    \label{fig:kvsp}
\end{figure}

We also tested the dependence of $k^*$ on the other parameters $p$, $v_{max}$ and $t_{tl}$. In particular, there is no explicit dependence on $p$ (only via $\rho_r$) and no dependence on $v_{\max}$ and $t_{tl}$ (see Fig.~\ref{fig:kvsp}).

This leads us finally to the following scaling form for $k^*$
\begin{equation}
    k^* \approx \frac{\rho_r}{L_{car}} \left(\frac{\rho_r}{L_{car}\rho_i}\right)^{\beta}
    \label{eq:kstar}
\end{equation}
with $\beta$ given by Eq.~\ref{eq:beta}. The prefactor in this expression has a natural interpretation as it is the number of cars per unit area that can be fitted in the network. 

Fig.~\ref{data collapse}(b) shows that the scaling of $q^*$ is more complicated and displays essentially two regimes, and that, unlike $k^*$, $q^*$ also depends on $p$ and $t_{tl}$. We used a simple ansatz suggested by Fig.~\ref{data collapse} of the form $V^*(n,p,t_{tl}) = V_{lim}\left(1-\mathrm{e}^{-n/n_c(p,t_{tl})}\right)$
where $n_c(p,t_{tl})$ is a crossover value that depends on $p$ and $t_{tl}$ (but is also independent of $v_{max}$, see Appendix B), while the general exponential form of $V^*$ and the limit $V_{lim}$ at large $n$ remain are independent of $p$, $t_{tl}$ or $v_{\max}$. Using the data collapse (Fig.~\ref{data collapse}(b)), we estimate for $p=0$ the critical value $n_c \approx 20$ cars/road, and the limit for large $n$ gives us the value $V_{lim}\approx 400$ veh/h/lane. We then get the complete scaling for $q^*$ under the form
\begin{equation}
    q^* \approx \frac{\rho_r V_{lim}}{L_{car}}\left(1-\mathrm{e}^{-n/n_c(p,t_{tl})}\right)
\label{eq:qstar}
\end{equation}

\begin{figure}[h!]
    \centering
    \begin{minipage}{1.1\linewidth}
    \includegraphics[width=\linewidth]{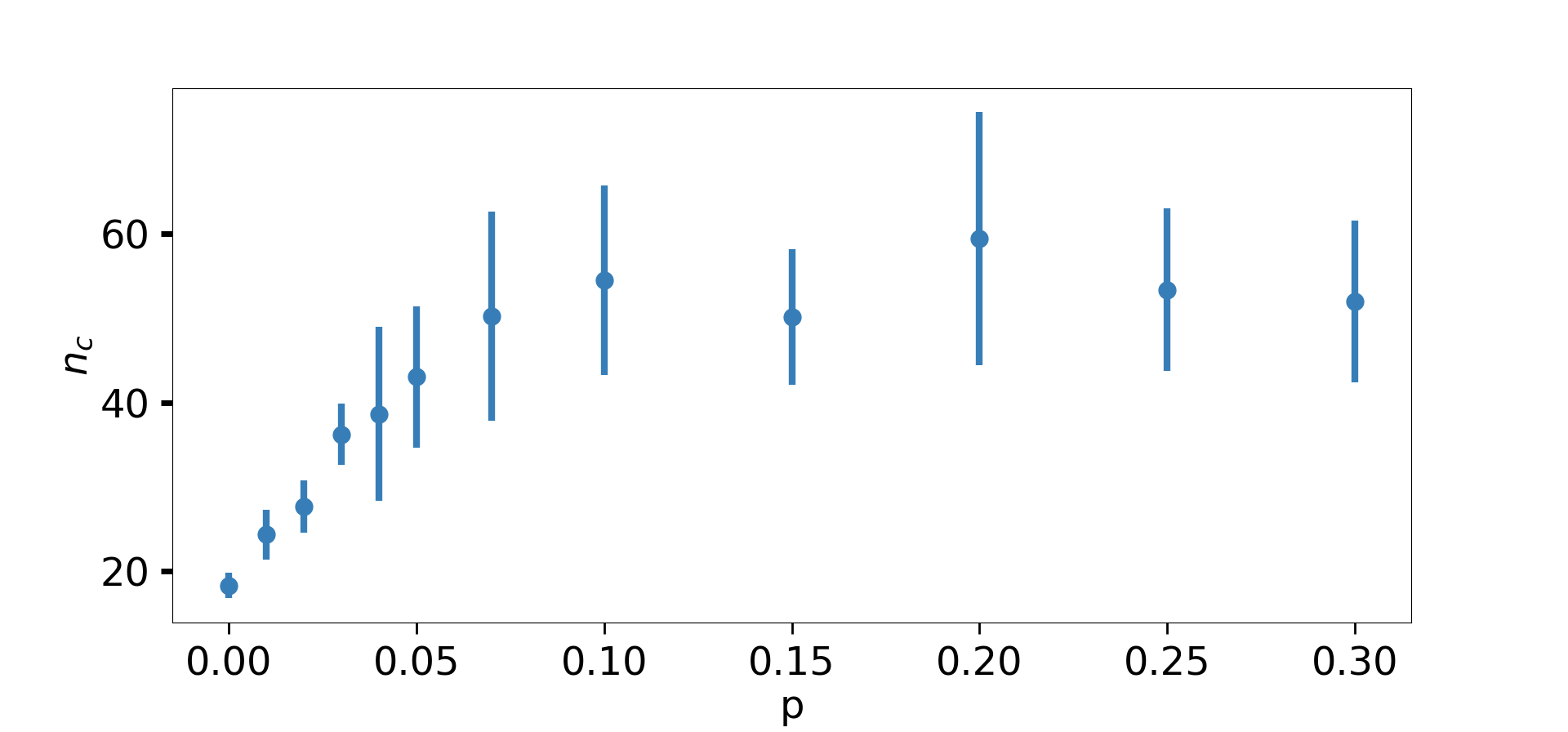}
    \end{minipage}
    \begin{minipage}{1.1\linewidth}
    \includegraphics[width=\linewidth]{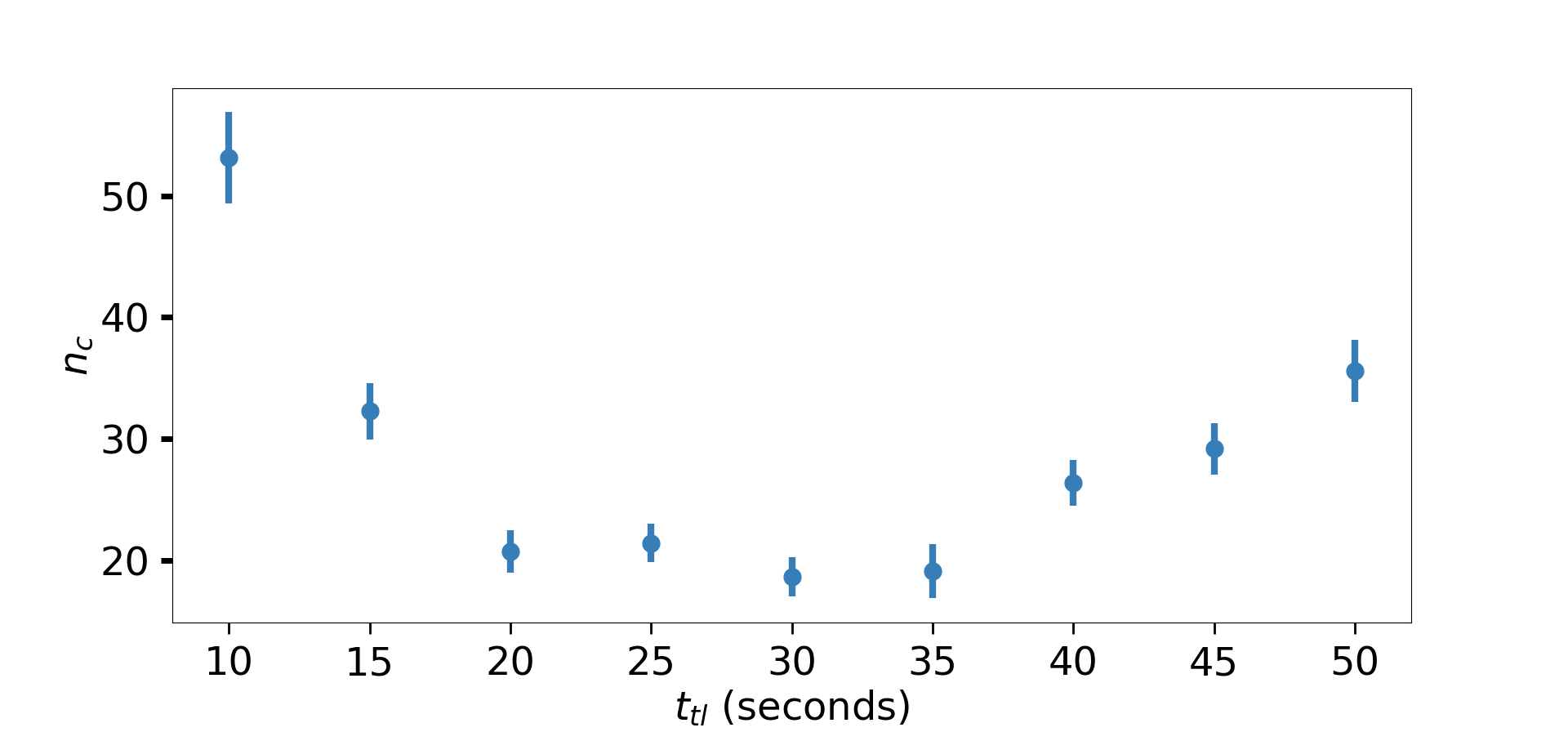}
    \end{minipage}
    \caption{Variation of $n_c$ (top) with $p$ (at fixed $t_{tl}=30$ seconds) and (bottom) with $t_{tl}$ (at fixed $p=0$).}
    \label{fig:n_c}
    \vspace{-0.3cm}
\end{figure}

The shape of the function $V^*$ and the influence of $p$ and $t_{tl}$ deserve some discussion. The quantity $n_c$, that characterizes the maximum number of cars that can be fitted on the network, separates two regimes for the function $V^*$. For $n>n_c$, $V^*$ is essentially constant, meaning that $q^*$ becomes proportional to $\rho_r$ and does not depend on $\rho_i$. If the roads of the network are long enough, the MFD will then be governed by the roads themselves and the impact of intersections will be negligible. This also gives an interpretation for $V_{lim}$ which is the average speed of a car travelling on the empty network but respecting all traffic lights. On the opposite side of the spectrum, if roads are short ($n<n_c$), the MFD will be mostly determined by intersections. Two mechanisms are at play here. On one hand, cars will spend a larger fraction of their time accelerating or decelerating, which is suboptimal, and on the other, short roads increase the likelihood of a spillback of congestion into an intersection, leading to a gridlocked network leading to a decrease of $q^*$. This is consistent with the observed increase of $n_c$ with $p$ shown in Fig.~\ref{fig:n_c}. 
For $p=0$, the network is completely regular and there is no reason for the traffic to accumulate at a given place. The probability of a gridlocked intersection is small and even relatively short roads are sufficient to prevent spillbacks. However, if one removes some links ($p>0$), breaking this regularity, one introduces congestion hotspots into the network. More specifically, the absence of some links creates intersections where two roads merge into one. In order to be resilient against spillbacks, roads of the resulting network have to be at least twice as long, which is consistent with the observed increase of $n_c$ from $\approx 20$ to $n_c \approx 50$ for $p>0$. We notice here that the transition occurs in the range $p\in [0;0.1]$. For larger values of $p$, $n_c$ remains constant, which is consistent with the proposed mechanism. Note that this discussion implies that the global efficiency of the network, of which $q^*$ is a measure, is ultimately defined by the few (local) weakest points of the network. Creating only few hotspots of congestion is all it takes to reduce the macroscopic capacity of the network by up to a factor 3. 

Note that this discussion is true only in the specific case of an homogeneous origin-destination matrix, studied here. In particular, having heterogeneous turning ratios at the intersections might change the results and is certainly something that should be studied in the future (see discussion). 

The variations of $n_c$ with $t_{tl}$, displayed in Fig.~\ref{fig:n_c} are consistent with previous results such as those obtained in the original study by Chowdury and Schadschneider \cite{ChowdhuryS99}. In the case where traffic lights are synchronized (green wave), the traffic flow on the network is maximized for a correctly chosen value of the traffic light duration. With our variables, this translates into a minimum for $n_c(t_{tl})$, reached for $t_{tl}\approx 30$ seconds. The reason for this optimum is a competition between two optimization problems. On one hand, cars need to accelerate from a standstill. Increasing $t_{tl}$ increases the average speed of cars passing the intersection during the green phase, thus increasing the flow. If $t_{tl}$ is too large, one reaches a point where few cars pass the intersection and a large number of cars wait at the red light of the intersecting road. This discussion is well known and can be found in more details in traffic engineering handbooks such as \cite{NCHRP}. Our optimal value of $t_{tl}\approx 30$ seconds is in good agreement with green light duration chosen in typical urban networks, where synchronization strategies for traffic lights are also commonly implemented.

\section{Comparison with empirical results}

\begin{figure}
    \centering
    \includegraphics[width=1.1\linewidth]{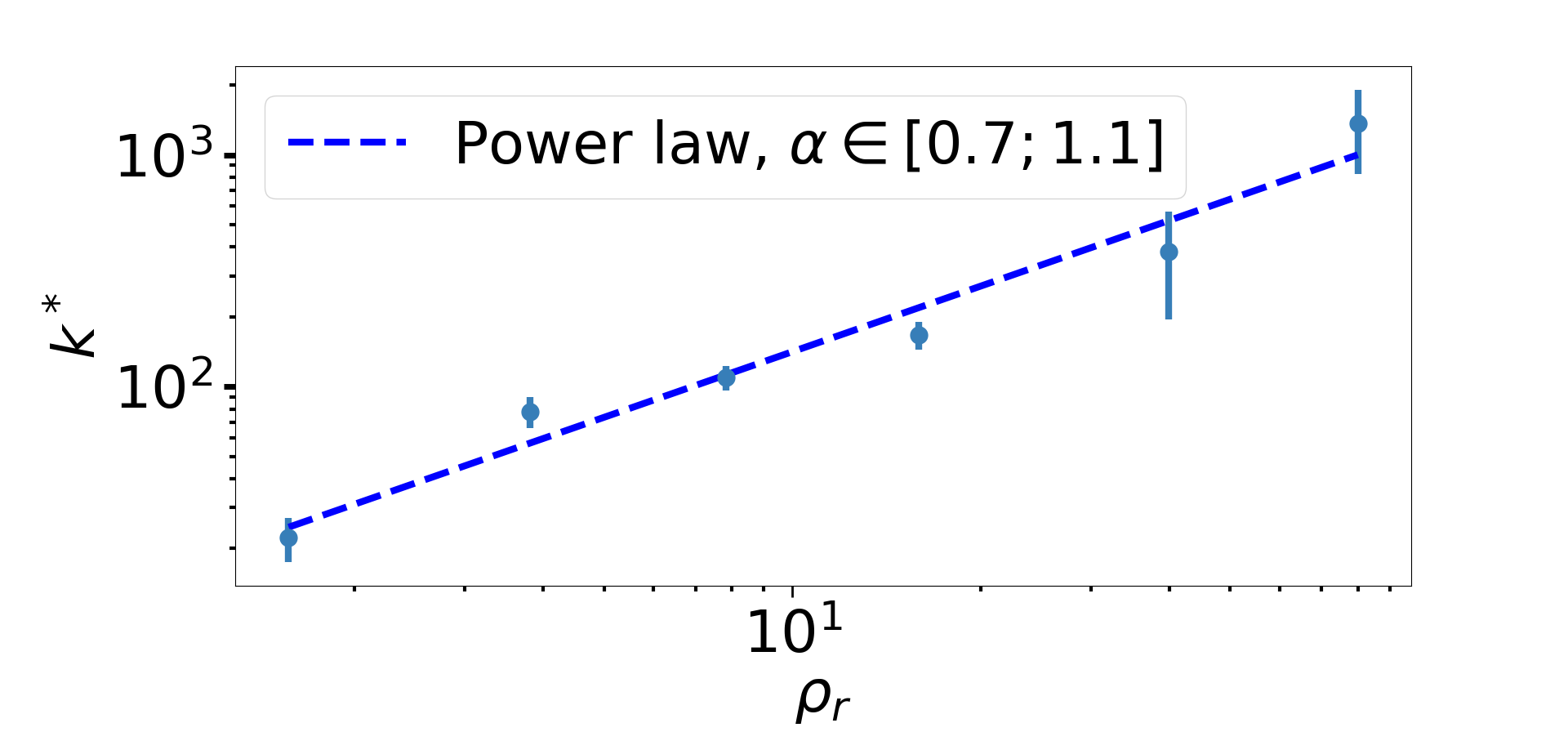}
    \caption{Influence of $\rho_r$ on $k^*$. A power law fit gives an exponent in $[0.7,1.1]$ in agreement with the scaling form result $k^*\sim \rho_r^{1+\beta}$ (network of $13\times 13$ nodes with $p=0$)}
    \label{fig:relations}
    \vspace{-0.3cm}
\end{figure}

Our scaling laws Eqs~(\ref{eq:kstar}, \ref{eq:qstar}) explain many results observed on MFDs, including the empirical relation found in \cite{LoderAMA19} between $\rho_r$ and $k^*$. In this study, they found a power law of the form $k^*\approx \rho_r^{\alpha}$ with exponent $\alpha=0.85$ and rejected the assumption of a linear relation between these variables. Our scaling form Eq.~\ref{eq:kstar} implies that
\begin{align}
    k^*\sim \rho_r^{1+\beta}
\end{align}
where $1+\beta\approx 0.9 \pm 0.2$ in agreement with the results obtained in \cite{LoderAMA19}. In Fig.~\ref{fig:relations} we plot the numerical result for $k^*$ versus $\rho_r$ confirming our scaling result. 

\begin{figure}
    \centering
    \includegraphics[width=\linewidth]{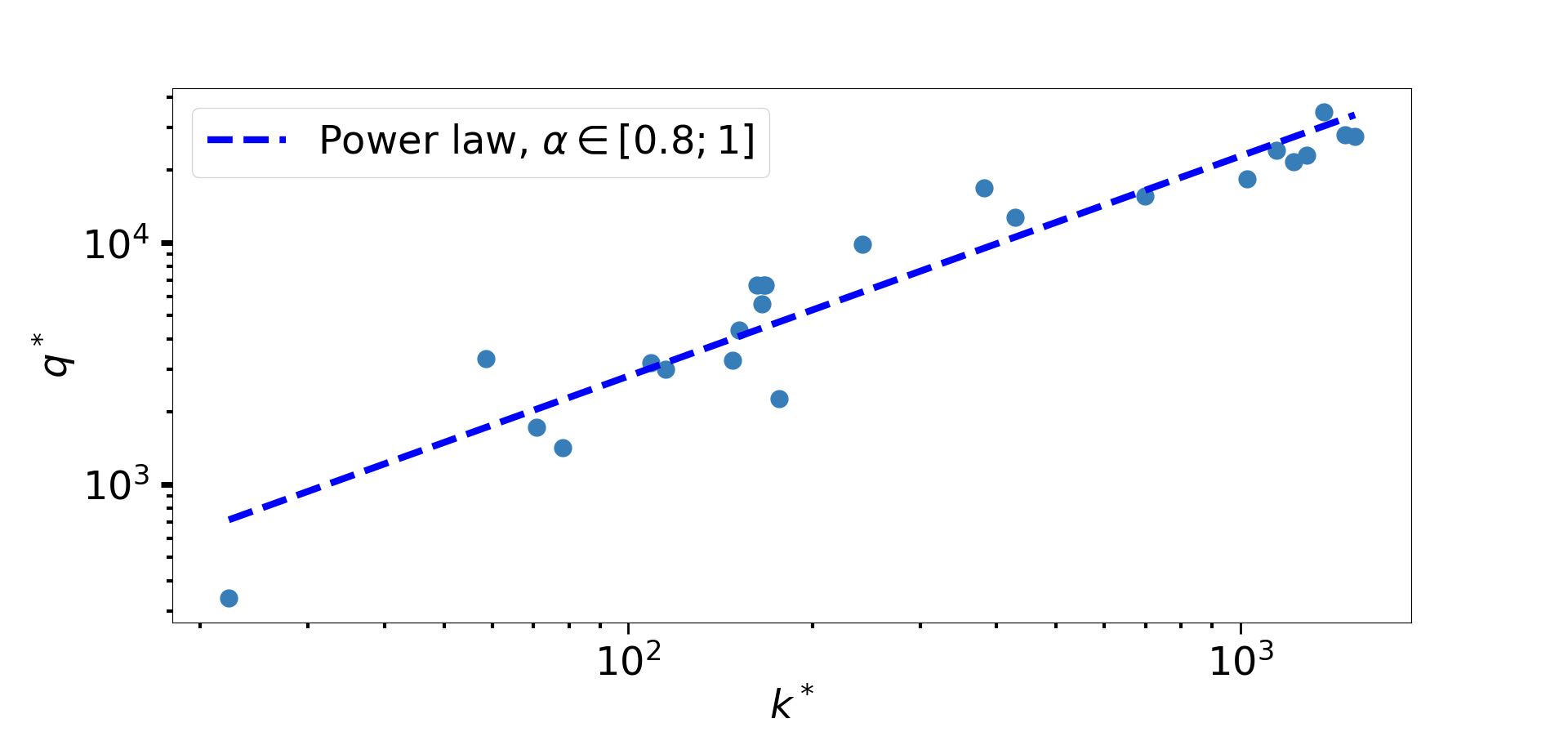}
    \caption{Simulation results showing the optimal flow $q^*$ as a function of $k^*$. The dashed line is a power law fit with exponent  $0.9 \pm 0.1 \; (R^2 = 0.9)$.}
    \label{fig:q_k}
\end{figure}
\begin{figure}
    \includegraphics[width=\linewidth]{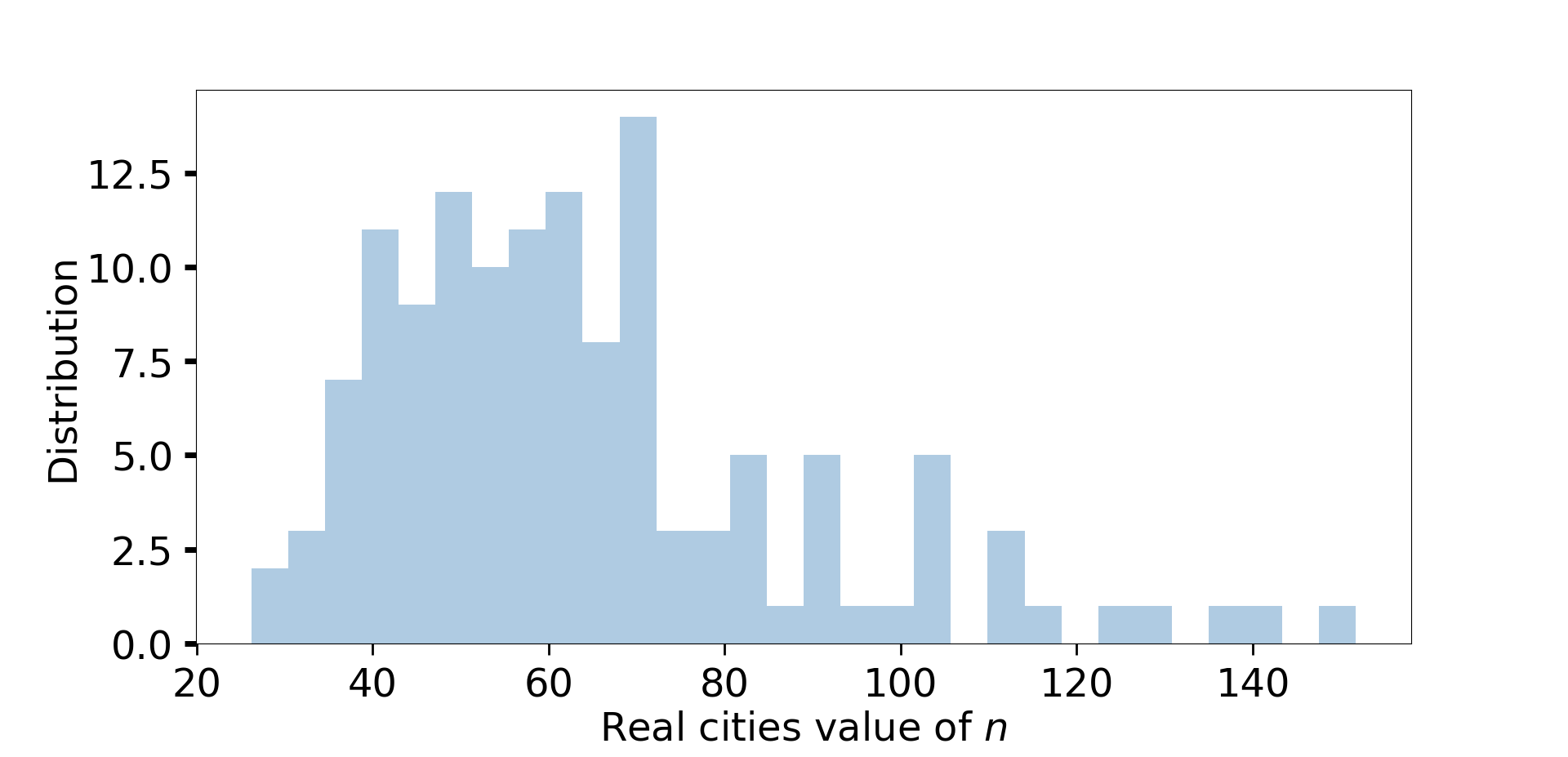}\\
    \includegraphics[width=\linewidth]{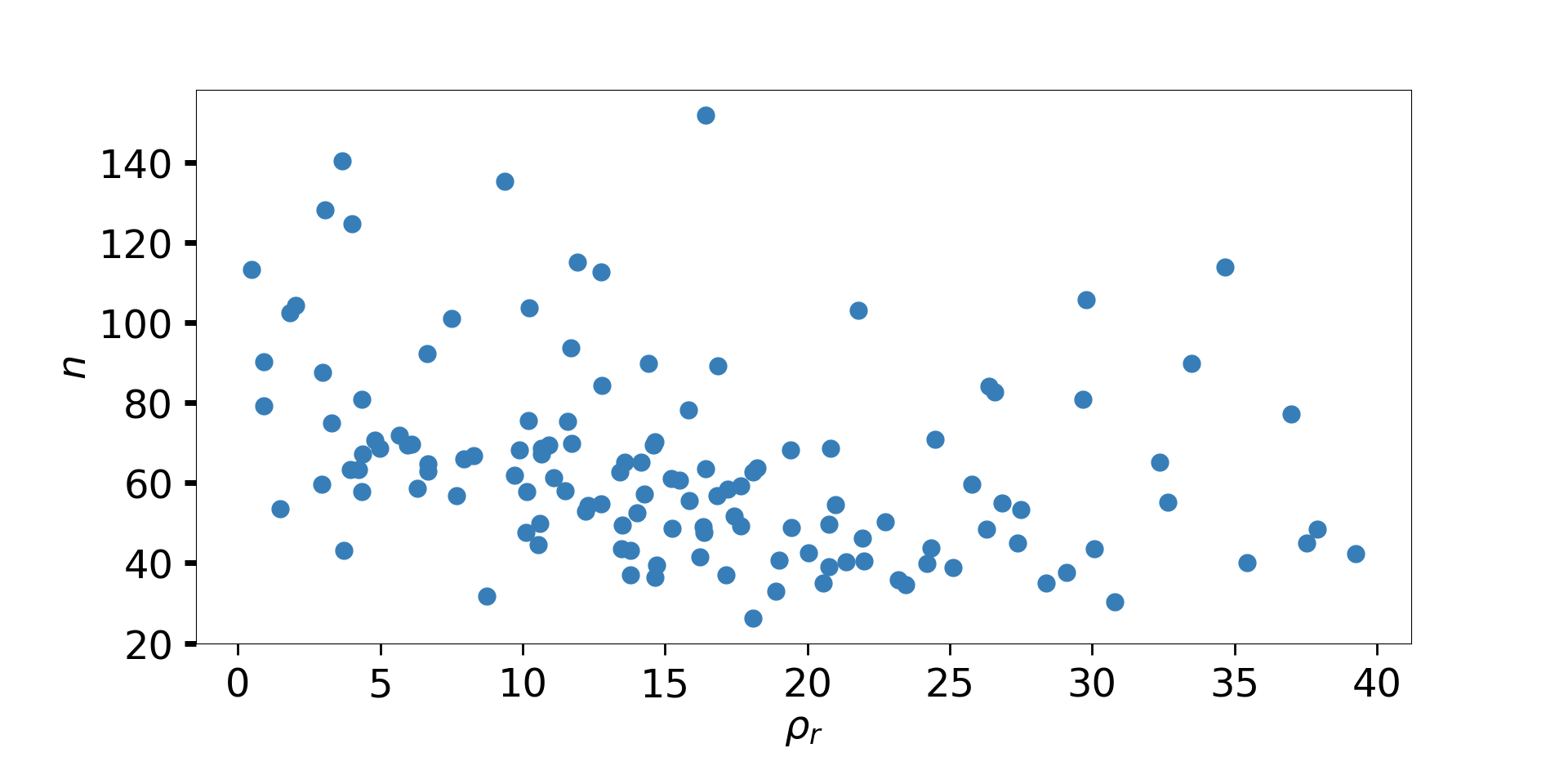}
    \caption{Top: Distribution of the value of $n$ in real cities, varying by not more than a ratio 5. Bottom: Influence of $\rho_r$ on the value of $n$ for real cities, showing close to no correlation ($r=-0.32$). Figures produced with \textit{openstreetmap} data, on the cities used for the empirical study \cite{LoderAMA19}.}
    \label{fig:distribution_n}
  \end{figure}

Loder et al. \cite{LoderAMA19} also found empirically a linear relationship between $k^*$ and $q^*$.
Numerically, we find (see Fig.~\ref{fig:q_k}) an almost linear relation between both parameters, $q^* \approx (k^*)^{\alpha}$ with $\alpha = 0.9 \pm 0.1 \; (R^2 = 0.9)$, close to what Loder et al. found.
Our scalings however show that this linear relationship is not true \textit{a priori}. In fact, using these scaling forms, the ratio $q^*/k^*$ can be rewritten as
\begin{equation}
    \frac{q^*}{k^*} =
    V_{\lim}\; n^{-\beta}\; \left(1-\mathrm{e}^{-n/n_c(p,t_{tl})}\right)
\label{ratio}
\end{equation}
where we recall that $n=\rho_r/(L_{car}\rho_i)$. For real networks, the variations of $\rho_r, \rho_i, L_{car}$ are such that $n$ does not vary over more than a decade (typically not more than a factor of order $6$), and is usually in the range $[25,100]$ (see Fig.~\ref{fig:distribution_n}(a)). This leads to $1-\mathrm{e}^{-n/n_c}$ varying by a factor $\pm 2$, while the variations of $n^{0.1}$ are negligible. $V_{lim}$ is also known to vary between cities, but again by not more than a factor $1.5$ (see for instance \cite{LoderAMA19}). Note that fig.~\ref{fig:distribution_n}(b) shows that, interestingly, the average length of roads does not seem to be a function of road density in real cities.

These empirical considerations thus show that both quantities $k^*$ and $q^*$ can vary independently between cities by up to a factor $\pm3$. However, as shown in \cite{LoderAMA19}, both quantities vary by more than a factor $10$ mainly under the common influence of $\rho_r$. This common $\rho_r$ term makes them seem proportional, even though it is not strictly speaking the case (as illustrated in figure \ref{fig:q_k}). We also note that a simple approximation for computing the MFD, as used in \cite{KnoopLH15}, further suggests that $k^*$ and $q^*$ have no reason to be proportional.

Rather counter-intuitively, our results show a limited influence of the intersection density on the MFD. More precisely, $\rho_i$ has no influence on $k^*$, and only an indirect influence on $q^*$ through the length of the roads $n$ and its impact on $V^*$. This is consistent with \cite{LoderAMA19}, where only a slight negative correlation between $\rho_i$ and $q^*$ was observed. Another empirical result in \cite{LoderAMA19} is the reduction of the network capacity when the bus production increases. Our scaling for $q^*$ indeed comprises the length of vehicles and by adjusting the average length of vehicles according to the proportion of buses, we reproduce exactly the empirical results in \cite{LoderAMA19}, giving another confirmation to our proposed scaling laws (details and calculations in Appendix C). Our results thus confirm and refine empirical findings observed about the MFD and provide a framework under the form of scaling laws for describing its critical properties.

\section{Discussion}

\begin{figure}[b!]
    \includegraphics[width=1\linewidth]{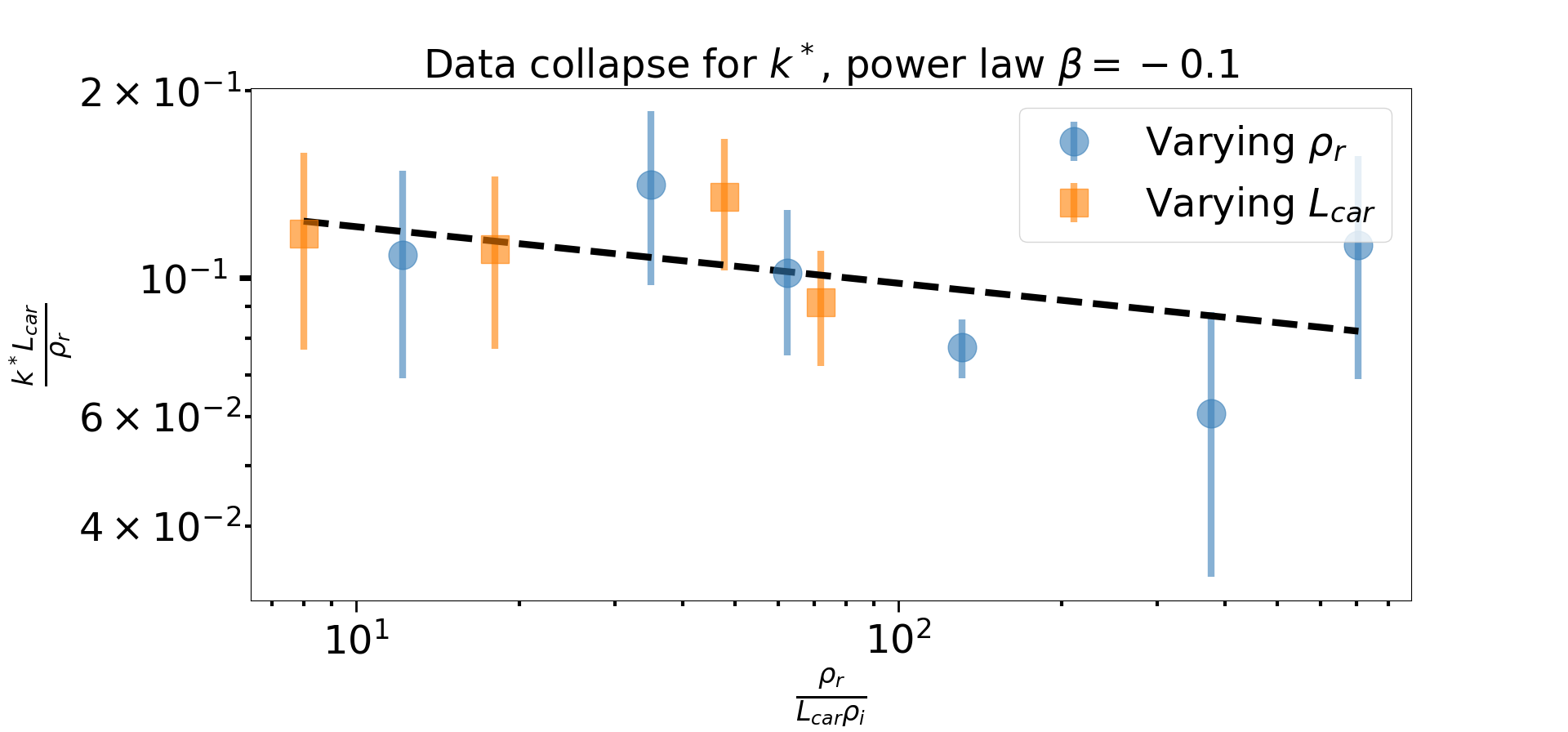}\\
    \includegraphics[width=\linewidth]{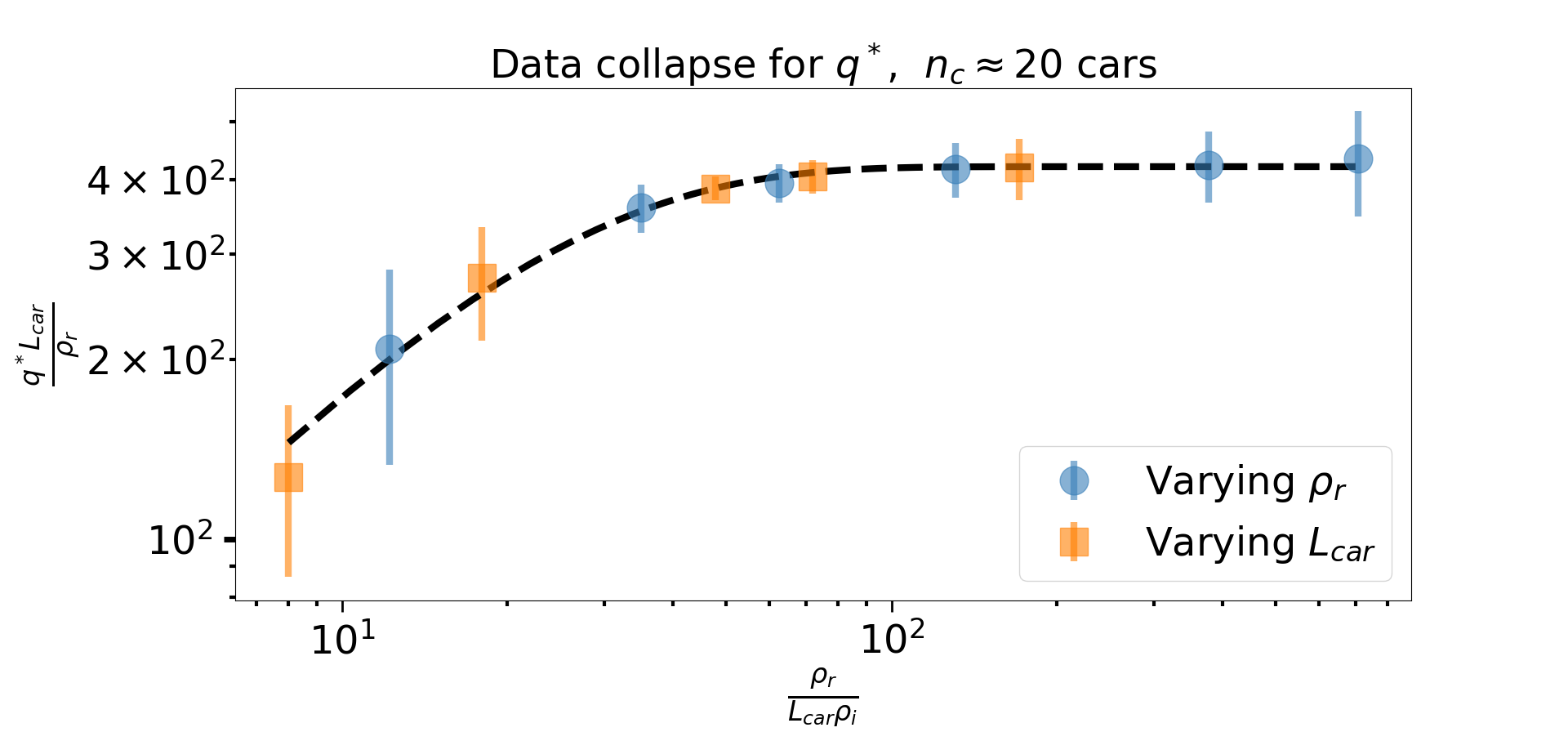}
    \caption{Data collapse for $k^*$ (left) and $q^*$ (right) in the case of a network with variable road lengths, with $p=0$. The average behavior is the same as for the regular network described in the article, but the errorbars are larger, indicating that the position of the roads becomes relevant.}
    \label{fig:collapse_variable_length}
\end{figure}

There are many interesting directions for future theoretical research about the MFD. First, our model considered here a specific origin-destination matrix (with all cars going from west to east and south to north) and the good agreement with the empirical measures of \cite{LoderAMA19} suggests that this is possibly irrelevant. Also, note that we studied a regular lattice with missing links, but more generally, we could think of more disordered networks, with different types of roads (so far in our simulation, all roads are identical), or different topologies. Preliminary results obtained when roads have variable
lengths are shown in Fig.~\ref{fig:collapse_variable_length}.
These results suggest that on average, the behavior of $k^*$ and $q^*$ with respect to the network parameters remains preserved. There are however large fluctuations of the MFD from one network to another when roads have variable lengths. When roads are different, we can indeed expect that their spatial distribution is relevant, leading to large fluctuations, but further studies are needed in order to confirm this.

We presented a simple model, mainly motivated by the idea of finding the relevant parameters for the MFD and
allowing for a scaling analysis of the problem. Introducing too many parameters would have made this difficult
if not impossible. Directions for further studies could be to include various origin-destination patterns, with cars
choosing their trajectory along shortest paths. Indeed, as we mentioned, bottlenecks seem to play an
important role in the MFD, and it is the interplay between network structure and OD-matrix that seems to be
relevant. Also, we focused on steady-state MFDs which allows to extract a `clean' MFD and to study
the influence of network parameters. However, transient regimes of loading and unloading are of prime
importance and occur more frequently than the ideal steady-state regime. Studying the network's response
during these phases would be of great interest but certainly requires additional parameters and a better
understanding of the MFD.

Finally, and from a more theoretical perspective, it is interesting to note that the concept of the MFD is similar to an effective medium theory \cite{zeng1988effective,clerc1990electrical,choy2015effective} relating macroscopic quantities (such as the flow and the density) of a composite system made of roads with different properties. More precisely, each road $e$ is described by a fundamental diagram of the form $q_e=f(k_e)$ (where $q_e$ and $k_e$ are the capacity and accumulation on the road $e$), and the MFD can be written under the form $q=\langle q_e\rangle=F(\langle k_e\rangle)$ (where the brackets denote the average over all roads). The problem is thus really to go from the microscopic (nonlinear) behavior to the average macroscopic behavior described by $q$ and $k$. The simplest approximation is to consider that $F$ is the average of the local function $f$ (which was done \cite{KnoopLH15}), but it would be interesting to go beyond this simple approximation and to develop an effective medium theory for urban traffic. In particular, this suggests the possible intriguing relation between fracture phenomena and congestion spread in networks.


%
    
\section*{Appendix A. Size effect}

\begin{figure}[!b]
\includegraphics[width=\linewidth]{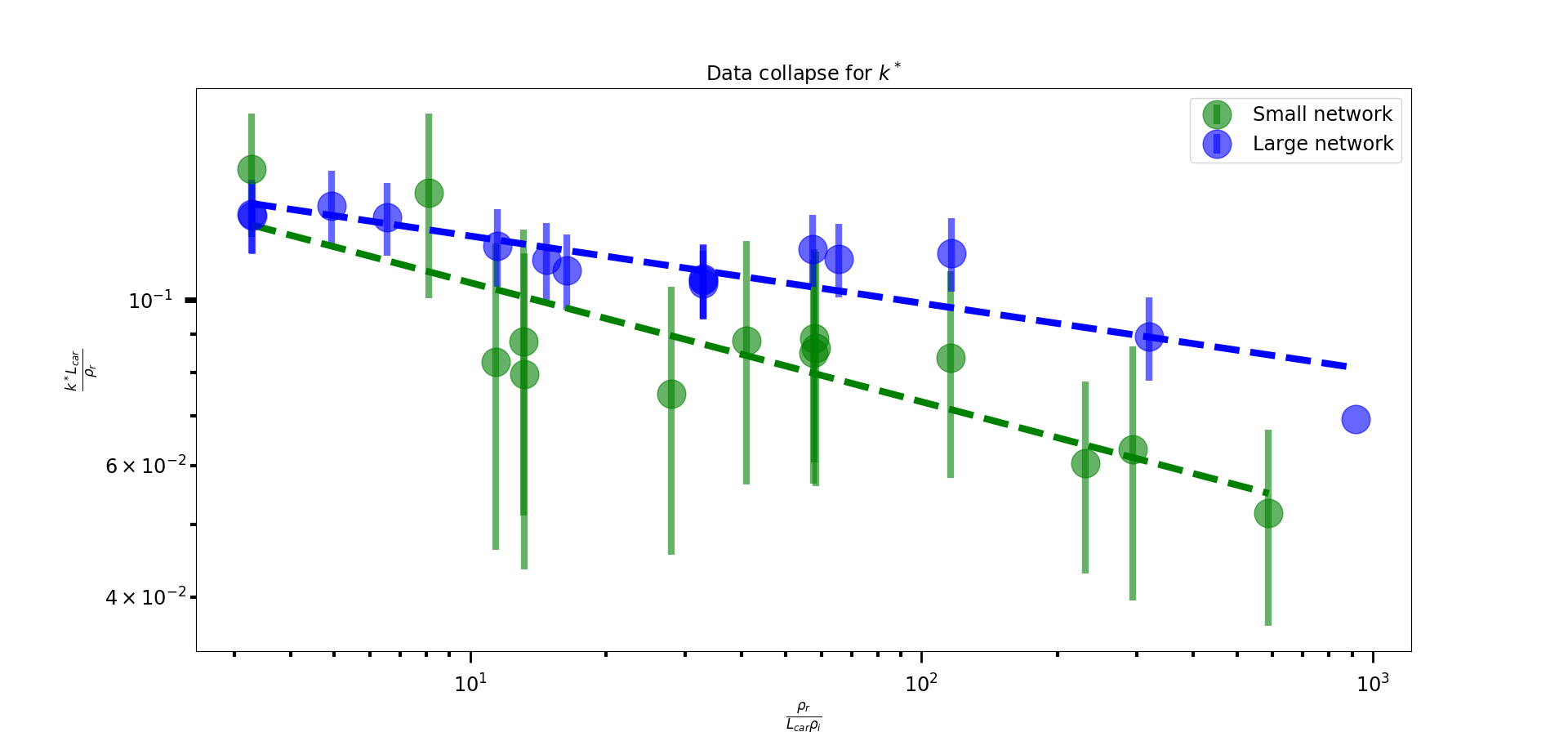}\\
\includegraphics[width=\linewidth]{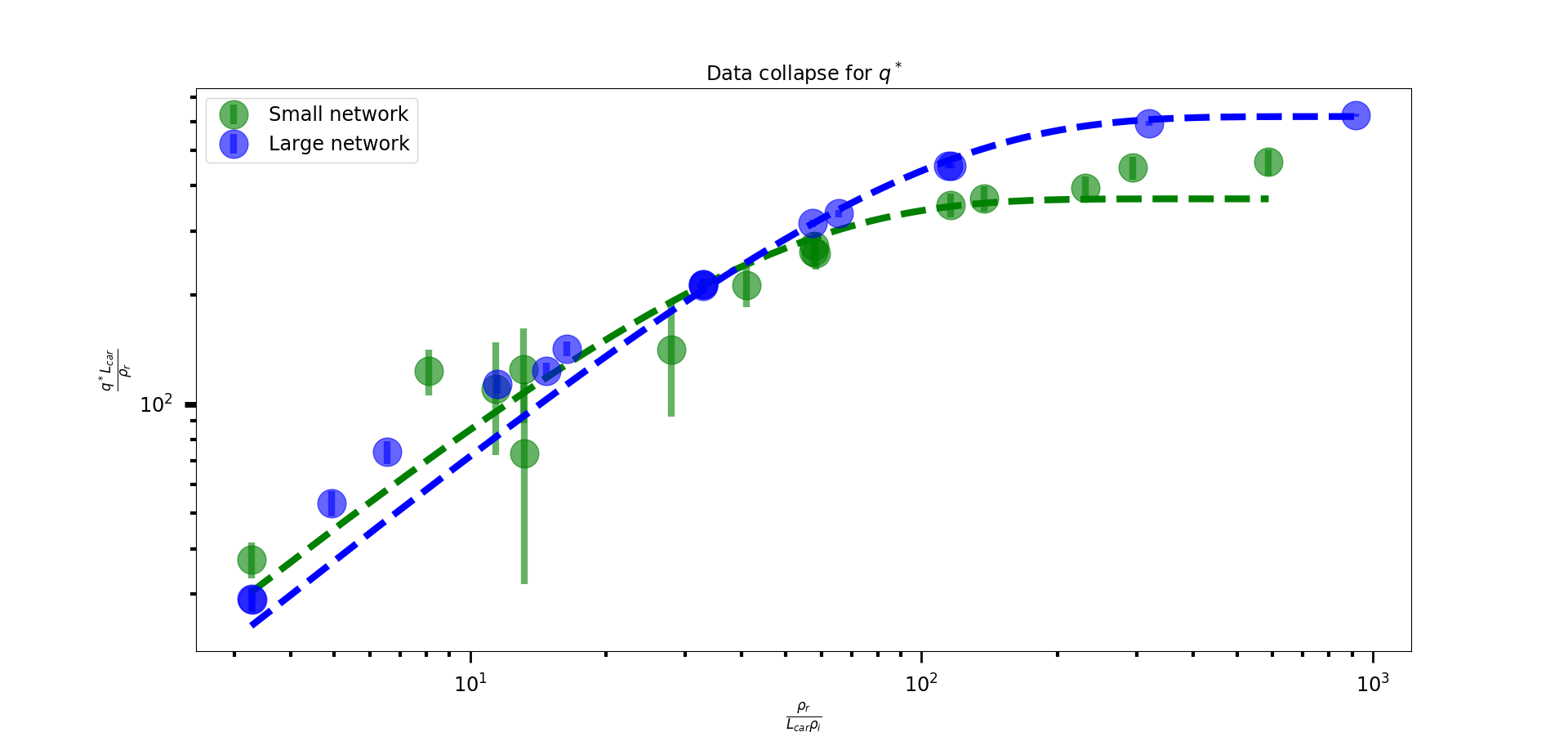}
\caption{Comparison of the data collapse in the case (small network: $13\times 13$, large network $20\times 20$). Top : for $k^*$, we find in both cases a power-law of very small exponent $\alpha \in [0 ; 0.2]$. Bottom: for $q^*$, we find in both cases an exponential fit, with a small difference, as now $n_c \in [15 ; 60]$ cars for the small network versus $n_c \in [70 ; 90]$ cars for the large network.}
\label{Compare_collapses}
\end{figure}

We show in Fig.~\ref{Compare_collapses}  the comparison of the data collapse in the case $p=0.2$, $t_{tl}=30s$ on a the $13\times13$ network and a $20\times20$ network.
We see on this figure that our results are stable with a small impact of the system size.

\section*{Appendix B. Influence of parameters on $q^*$}

\begin{figure}
    \centering
    \begin{minipage}{.5\linewidth}
    \includegraphics[width=\linewidth]{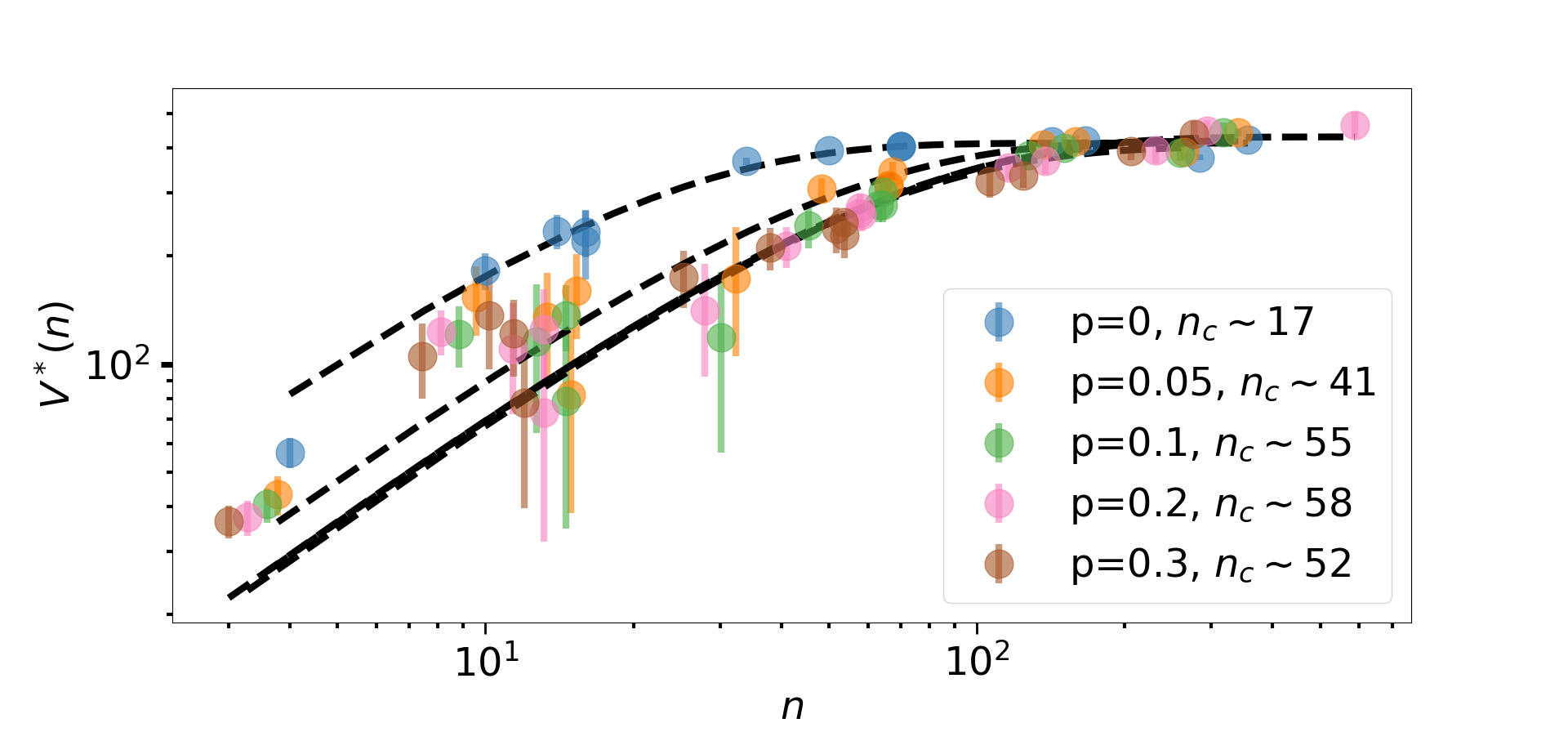}
    \end{minipage}%
    \begin{minipage}{.5\linewidth}
    \includegraphics[width=\linewidth]{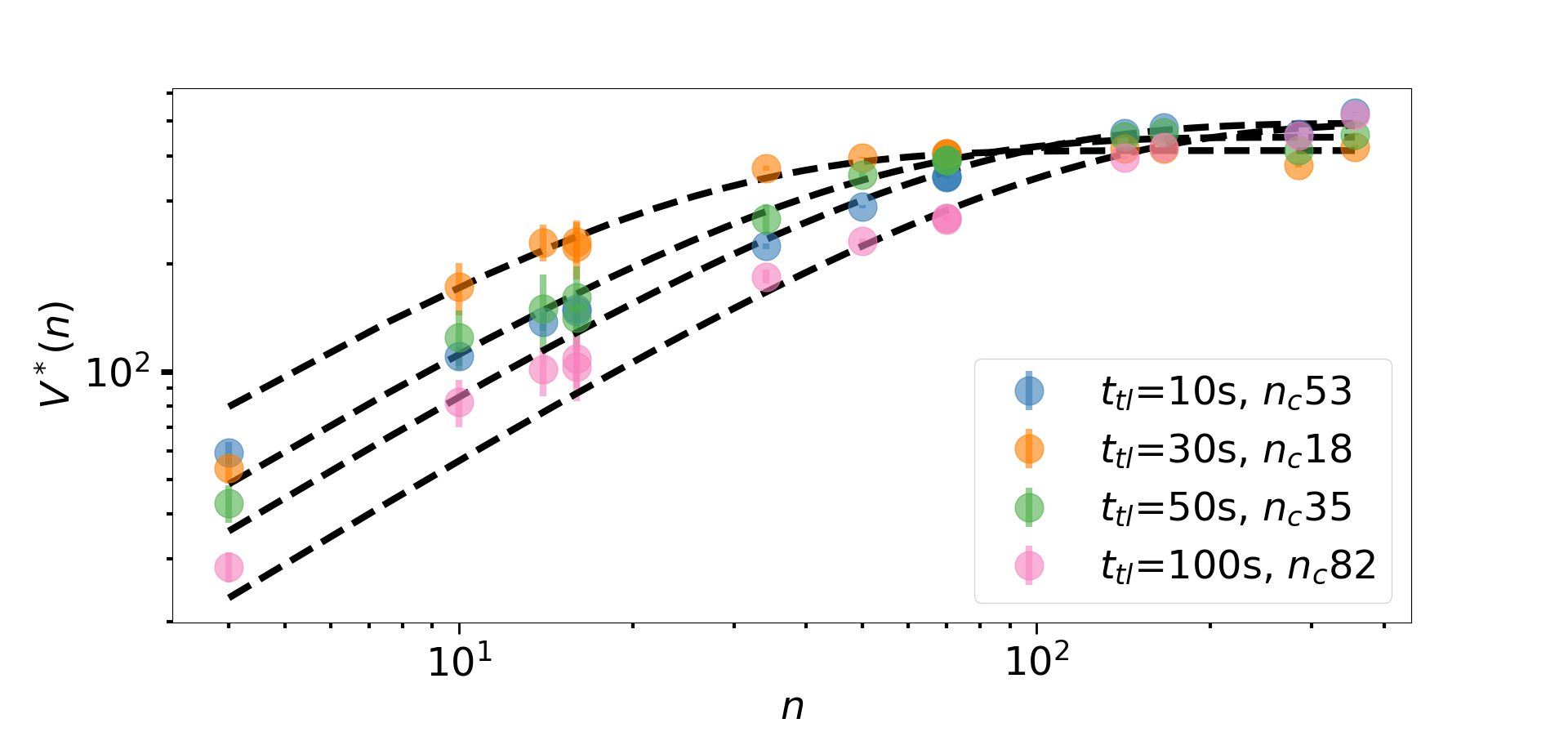}
    \end{minipage}
    \begin{minipage}{.5\linewidth}
    \includegraphics[width=\linewidth]{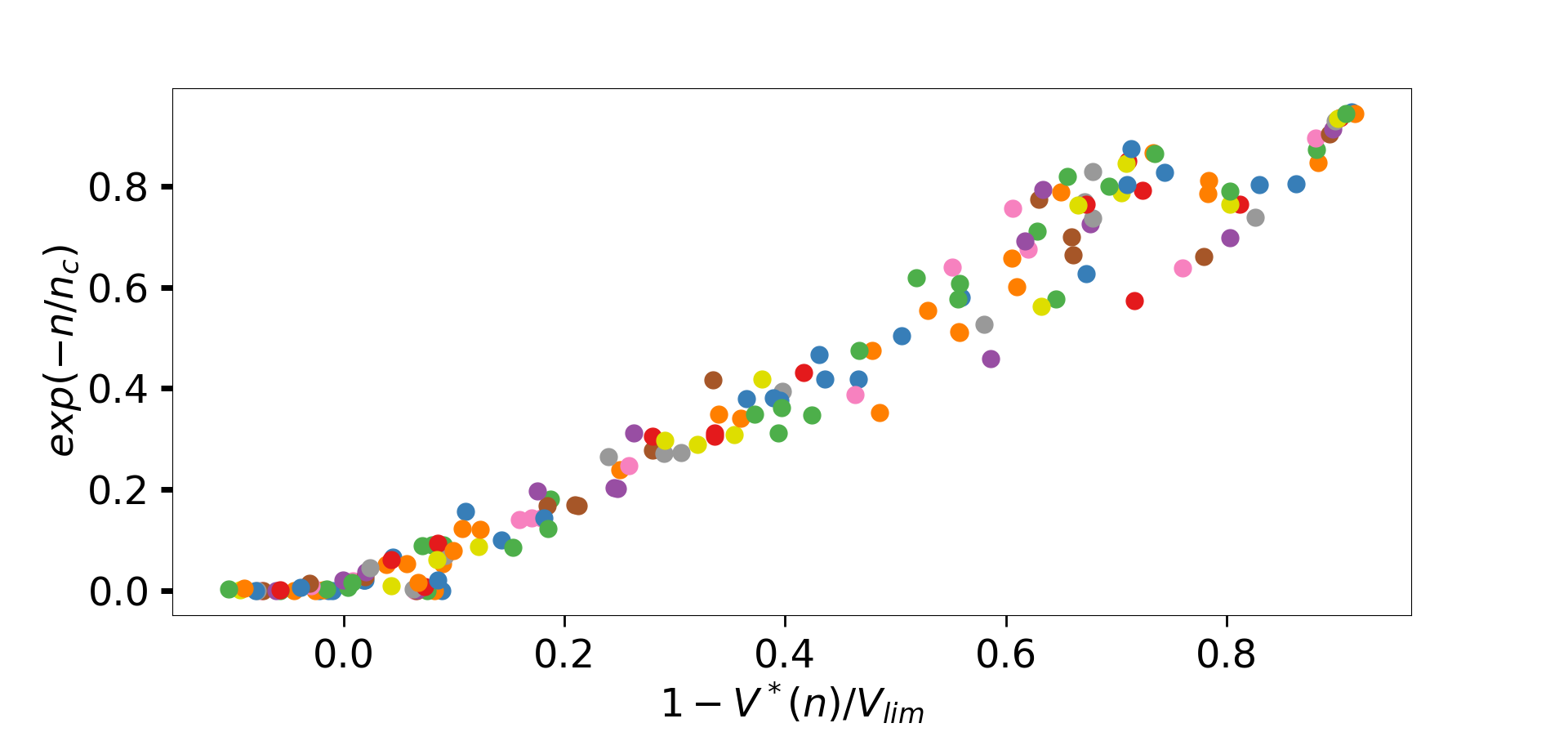}
    \end{minipage}%
    \begin{minipage}{.5\linewidth}
    \includegraphics[width=\linewidth]{./fig_dependence_p_V.png}
    \end{minipage}%
    \caption{Top: Data collapse for $q^*$ for different values of $p$ (at fixed $t_{tl}=30$ seconds) and $t_{tl}$ (at fixed $p=0$) and corresponding value of $n_c$. All fits have $R^2>0.98$. Bottom: Collapse of the dependence of $V^*$ to $p$ and $t_{tl}$ (same conditions). Each color corresponds to a value of $p$ (resp. $t_{tl}$)}
    \label{fig:qvsp}
\end{figure}

In Fig.~\ref{fig:qvsp}, we show the data collapse for $q^*$ for different values of $p$ and $t_{tl}$.

We verify that the dependence of $V^*$ to $p$ and $t_{tl}$ is indeed only through $n_c$ by plotting $\mathrm{e}^{-n/n_c(p,t_{tl})}$ as a function of $1-\frac{V^*}{V_{lim}}$ for various $p$ and $t_{tl}$ and verify that the graphs collapse into one (Fig.~\ref{fig:qvsp} bottom).

\section*{Appendix C. Influence of the bus production}
Another important result found empirically is that the capacity of the network is reduced when the bus production increases. More precisely, if we denote by $q^*(P_b)$ the maximum capacity when the bus flow is $P_b$ ($q^*(0)$ is then the capacity for the network when no bus is present), the capacity reduction of the network is defined by
\begin{equation}
\delta= \frac{q^*(0) - q^*(P_b)}{q^*(0)}\,.
\label{eq:capa}
\end{equation}
They express the bus production as a flux per unit surface, i.e. in the same units as the capacity. It is therefore natural to denote by $\theta_{b}$ the fraction of vehicles that are buses and we can then write $\theta_{b}=P_b/q^*(P_b)$. Using our proposed scaling, the main dependence of $q^*$ on the vehicle length is $q^*\sim 1/\; \overline{L_{veh}}$. If we consider that the length of a bus is $L_{b} \approx 3 L_{cars}$, the average length of the vehicles in the network is $\overline{L_{veh}} = L_{car}(2\theta_{b}+1)$, and we then obtain 
\begin{align}
\nonumber
 q^*(P_b) &= \frac{q^*(0)}{1+2\theta_b}\\
 &\approx q^*(0)-2P_b
 \end{align}
 which leads to a capacity reduction $\delta=2 P_b/q^*(0)$. We thus expect a linear relationship between the capacity reduction and the bus production, with a slope $2/q^*(0) \approx 10^{-4}\; km.h$ (considering the observed values for $q^* \approx 10^4 \; veh\cdot km/h/km^2$). This is in excellent agreement with the empirical results observed in \cite{LoderAMA19} which gives a slope $\approx 4. 10^{-4} \; km\cdot h$, and also explains the large dispersion observed in their results: for each value of $P_b$, the observed capacity reduction can vary significantly based on the capacity $q^*(0)$ of that particular city.



\bibliographystyle{apsrev4-1} 
\bibliography{refs}

\end{document}